\documentclass[preprint,11pt,authoryear]{elsarticle}

\makeatletter
\def\ps@pprintTitle{%
	\let\@oddhead\@empty
	\let\@evenhead\@empty
	\def\@oddfoot{\reset@font\hfil\thepage\hfil}
	\let\@evenfoot\@oddfoot
}
\makeatother

\usepackage{natbib}
\bibpunct[, ]{(}{)}{,}{a}{}{,}
\def\bibfont{\small}

\usepackage{graphicx}
\usepackage{rotating}
\usepackage{graphicx}
\usepackage{booktabs}
\usepackage{amsmath}
\usepackage{amssymb}
\usepackage{tikz}
\usetikzlibrary{shapes}
\usetikzlibrary{patterns}
\usetikzlibrary{arrows,positioning,decorations.pathmorphing,trees}
\usetikzlibrary{arrows.meta}
\usepackage{tikz-cd}
\usepackage{mathtools}
\usepackage{stackengine}
\usepackage{pgfplots}
\usepackage{float}
\usepackage{braket}
\usepackage{eurosym}
\usepackage{hyperref}
\usepackage{comment}
\usepackage{subcaption}
\usepackage{ifthen}
\usepackage{caption}
\usepackage{appendix}
\usepackage[format=default,font=footnotesize,labelfont=bf]{caption}
\usepackage{listings} 
\usepackage{color}
\usepackage{algpseudocode}
\usepackage{algorithm} 
\usepackage{float}
\usepackage{xurl}
\usepackage{array}
\usepackage{multirow}
\usepackage{makecell}
\usepackage{bm}
\usepackage{longtable}
\usepackage{lipsum}
\usepackage{fullpage}
\usepackage{setspace}

\newboolean{includeHidden}
\setboolean{includeHidden}{false}

\newcommand{\hide}[1]{\ifthenelse{\boolean{includeHidden}}{{\tiny\textbf{HIDDEN:~}#1}}{}}

\newboolean{tablesInText}
\setboolean{tablesInText}{true}
\newboolean{figuresInText}
\setboolean{figuresInText}{true}
\newboolean{footnotesInBrackets}
\setboolean{footnotesInBrackets}{true}

\newtheorem{result}{Result}

\journal{}

\begin{document}

\begin{frontmatter}
	
	\title{Challenges in Finding Stable Price Zones in European Electricity Markets: Aiming to Square the Circle?}
	
	\author[1]{Teodora Dobos\corref{cor1}}
	\ead{dobos@cit.tum.de}
	\author[1]{Martin Bichler}
	\ead{bichler@cit.tum.de}
	\author[1]{Johannes Kn\"orr}
	\ead{knoerr@cit.tum.de}

	\cortext[cor1]{Corresponding author}

	\affiliation[1]{organization={School of Computation, Information and Technology, Technical University of Munich},
		addressline={Boltzmannstrasse 3}, 
		city={Garching},
		postcode={85748}, 
		country={Germany}}
	
	\begin{abstract}
		The European day-ahead electricity market is split into multiple bidding zones with a uniform price. The increase in renewables leads to a growing number of interventions in the generation of energy sources and increasing redispatch costs. To ensure efficient congestion management, the EU Commission mandated a Bidding Zone Review (BZR) to reevaluate the configuration of European bidding zones. An integral part of this process was a locational marginal pricing study. Based on these prices, alternative bidding zone configurations were proposed. These bidding zones shall be stable and robust over time. For Germany, four configurations were suggested. 
        We analyzed the proposed configurations considering different clustering algorithms and periods based on the publicly released data set in the context of the BZR, and found that the configurations do not reduce the price standard deviations within zones much, and the average prices across zones are similar. Other configurations identified based on clustering the prices lead to lower price variance but they are not geographically coherent. Independent of the clustering features and algorithms used, the resulting clusters are not stable over time. Interestingly, the effect of a split on prices would be low based on an analysis of the BZR data set.
	\end{abstract}

	\begin{keyword}
		Energy pricing \sep Price zones \sep Bidding Zone Review \sep Electricity markets \sep Energy policy
	\end{keyword}
	
\end{frontmatter}

\section{Introduction}
\begin{figure}[!htp]
	\centering
	\includegraphics{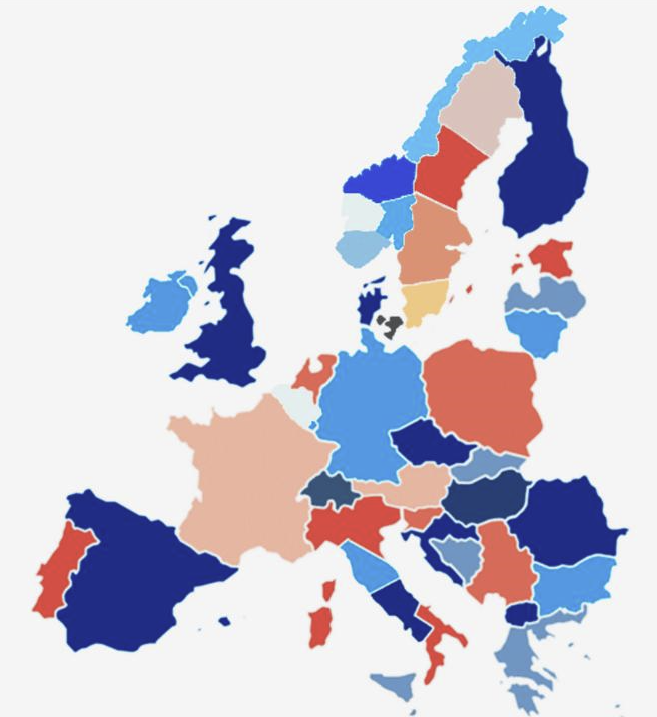}
	\caption{European Bidding Zones (adapted from \cite{florence-eu-bzs-figure})}
	\label{fig:European_bz}
\end{figure}

\noindent The European electricity market is split into multiple bidding zones (BZs) or price zones. Within these zones, market participants can submit bids and offers, which are subsequently matched without considering the capacities of the transmission network.\footnote{We will use the terms price zone and bidding zone interchangeably.}
\textcolor{black}{As shown in Figure \ref{fig:European_bz},} BZs are generally defined depending on national borders, such that a country represents a single bidding zone (e.g., France and Poland). However, there are also countries that are divided into several BZs, such as Norway, Sweden and Italy, or BZs that consist of more than one country, like the German-Luxembourgish bidding zone.

\color{black}
The implementation of European bidding zones began with the First Energy Package in the late 1990s, which aimed to liberalize national electricity markets and set the stage for cross-border trading \citep{meeus2020european-markets-evolution}. 
This was followed by the Second Energy Package in 2003, which further opened the markets and led to the organization of national electricity grids into bidding zones, often aligned with national borders. 
In 2006, the Nordic region pioneered market coupling, integrating Norway, Sweden, Finland, and Denmark into a regional electricity market with distinct bidding zones. 
The Third Energy Package of 2009 formalized the concept of cross-border bidding zones as part of creating a single European electricity market. 
Most recently, the introduction of Network Codes \citep{CR2013-network-codes} aimed to harmonize electricity market operations across Europe, enhancing cross-border coordination and the efficiency of bidding zones.
\color{black}

An important characteristic of \textcolor{black}{European} BZs is that in the day-ahead market, they implement a \textit{uniform pricing mechanism}, meaning that a single market-clearing electricity price applies to all participants belonging to a BZ.
\textcolor{black}{Europe's adoption of zonal pricing over the more granular locational marginal pricing (LMP) model seen in markets like PJM reflects a balance between economic efficiency, simplicity, and political practicality \citep{meeus2020european-markets-evolution, knoerr2024nodal-zonal,bjorndal2001}. While LMP offers greater efficiency by reflecting real-time grid conditions at each node, it introduces operational complexity and price volatility. In contrast, the European approach of implementing a}
uniform price in a whole BZ promotes transparency due to a single price signal in a large region, and it limits the market power that individual parties can exercise. \textcolor{black}{Moreover, this approach enhances liquidity and makes cross-border trading easier in a multi-country system \citep{nemo2020euphemia}.} 
However, a uniform electricity price also leads to significant negative ramifications. There are no local incentives to adapt to changes in the volatile supply of renewables, there are significant welfare losses due to the uniform price constraint, and large redispatch costs \textcolor{black}{to resolve internal congestion}. The latter is because transmission constraints are ignored within a BZ when supply is matched with demand \citep{german-grid-swamped}, and, \textcolor{black}{therefore,} the market-based allocation is not physically feasible. In addition, uniform prices do not set appropriate investment incentives. 

As an example, in the German-Luxembourgish BZ strong winds along the North Sea and Baltic Coast can contribute to elevated power production from wind farms in northern Germany.
The energy produced in the north must be transported to the demand centers in the south of the country. However, the transmission lines connecting these regions might become congested and thus electricity cannot be delivered as determined in the day-ahead market. To stabilize the grid, redispatch by the Transmission System Operators (TSOs) is required. TSOs often have to throttle the generation of northern conventional and wind power plants. At the same time, they instruct power plants in the south to increase energy production.
Such redispatch actions incur substantial costs. In 2023, the redispatch costs in Germany were 3.1 billion Euros.\footnote{\url{https://www.bundesnetzagentur.de/}}

\subsection{The Bidding Zone Review}
Given these considerations, in August 2022 the European Union Agency for the Cooperation of Energy Regulators (ACER) decided that alternative bidding zone configurations should be analyzed as part of the Bidding Zone Review (BZR), which is organized by the European Network of Transmission System Operators for Electricity \citep{bz-website}.
The goals of the BZR are described on a high level in different documents. 
According to Commission Regulation (EU) 2015/1222 and 2019/943, bidding zones ``should be defined in a manner to ensure efficient congestion management and overall market efficiency" \citep{CR2015-zones} and ``shall be based on long-term, structural congestions in the transmission network'' \citep{CR2019-zones}. 

Annex I of ACER's decision on alternative bidding zone configurations to be considered in the bidding zone review process \citep{acer-list-all-proposed-bzs} describes that different configurations were identified by solving a clustering problem ``\textit{based on nodal prices as clustering feature}''. The document argues that nodal prices ``\textit{can be used as proxies for economic efficiency, in line with the objectives of the Electricity Regulation}'', and that the objective of clustering algorithms is to ``\textit{minimize price dispersion within each bidding zone}.'' The document discusses three clustering algorithms: constrained K-Means clustering, Spectral Clustering with constrained K-Means, and Spectral Clustering with Constrained Deterministic Iterative Refinement Clustering (CDIRC). Moreover, a constraint related to the minimum size of each zone was included.\footnote{The threshold depends on the number of price zones under analysis and is between 7\% and 10\% of the nodes.}  

\begin{figure}[!htp]
	\centering
	\includegraphics[width=.6\textwidth]{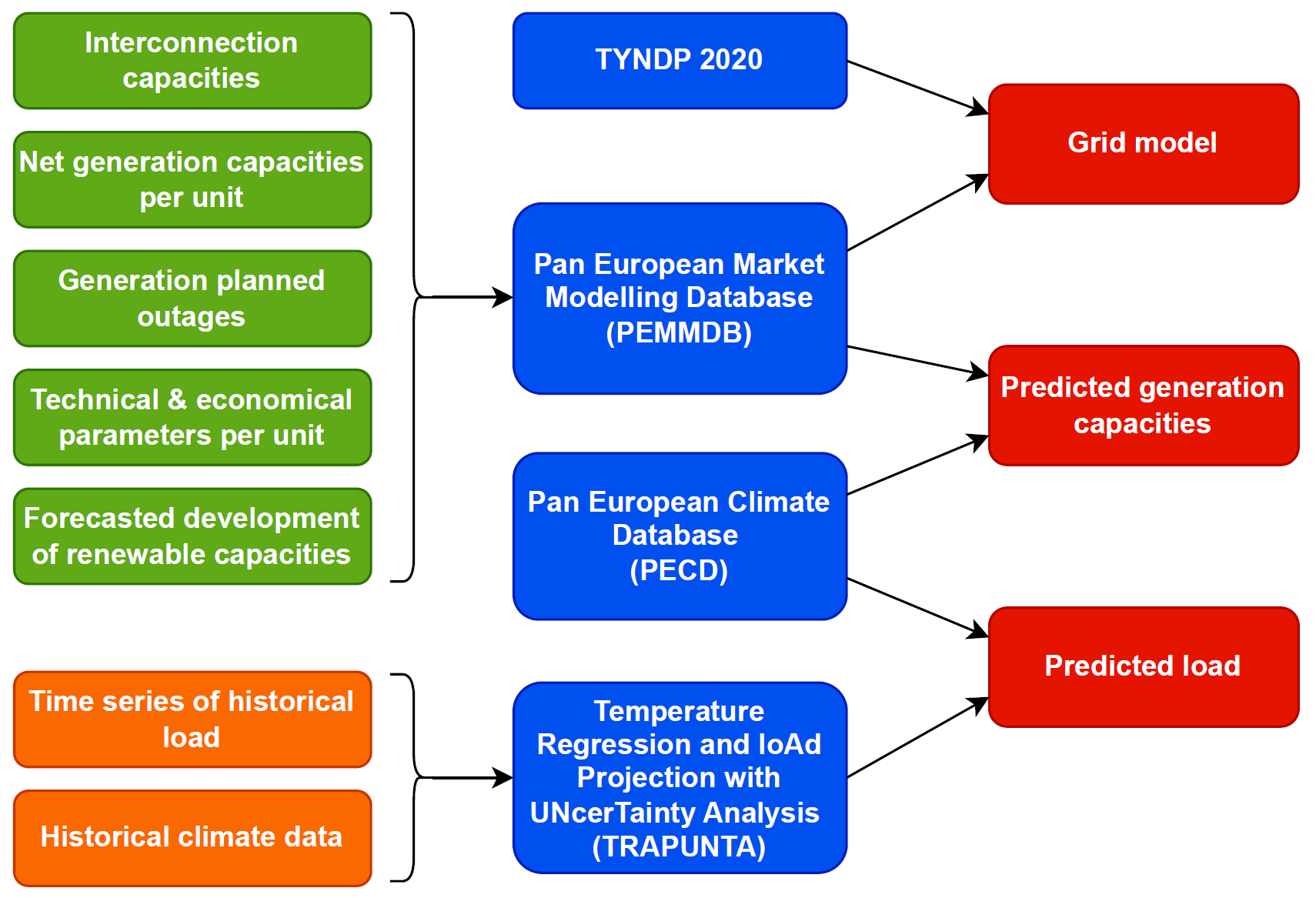}
	\caption{Overview of data sources and models used in the LMP Study}
	\label{fig:data_overview}
\end{figure}

Based on this ACER methodology \citep{acer-methodology}, ENTSO-E was tasked to calculate locational marginal prices (essentially, a nodal price for each substation in the transmission grid\footnote{https://www.tennet.eu/news/bidding-zone-review-tsos-investigate-alternative-bidding-zone-configurations}) for the target year 2025, for which generation and demand scenarios were created. \textcolor{black}{These scenarios originate from ENTSO-E's Data \& Models working group and were obtained following the procedure described \textcolor{black}{in Annex 1 of the LMP Study Report} \citep{bzr-report}. 
Weather data from} \textcolor{black}{the most representative combination of three years in terms of occurrence of residual loads\footnote{The residual load is defined as total demand minus renewable generation.} was used to create scenarios. 
This combination was identified considering the years between 1987 and 2016 by an algorithm that used the solar and wind infeed, as well as the hydro inflows and the load as variables.
The algorithm selected the combination for which the aggregate dataset of hourly residual loads has the highest average mean and standard deviation relative to all other combinations.}
\textcolor{black}{Eight representative weeks were identified for each year, with at least one week per season.}
\textcolor{black}{The grid model, the generation capacities and the load were predicted for 2025 based on several data sources and models which are shown in Figure \ref{fig:data_overview}.} 

\textcolor{black}{ENTSO-E computed bihourly nodal prices for the selected eight weeks in a report published in June 2022 \citep{bzr-report}. The prices were obtained based on a linear unit commitment model.}
As outlined above \citep{acer-clustering-algo}, the time series of nodal prices represented the clustering features that were used to partition the set of nodes in the transmission network into a predefined number of clusters or BZs. A BZ configuration or a clustering thus represents the assignment of every node to a specific cluster or price zone. 
Finally, ACER suggested four alternative bidding zone configurations for Germany \cite{acer-list-all-proposed-bzs} based on the results from the Locational Marginal Pricing (LMP) Study \citep{bzr-report, bz-website}. 
The evaluation of these configurations proposed by ACER in the BZR is ongoing at the time of writing.
In this process, the TSOs are evaluating 22 indicators related to \textit{market efficiency}, network security, and the \textit{stability and robustness} of the proposed configurations over time \citep{acer-methodology}.

\subsection{Contributions}
In this paper, we leverage the data from the LMP study provided by ENTSO-E and study how stable and robust the configurations of price zones (or, clusterings) proposed by ACER are with respect to different time frames, different inputs (i.e., nodal prices with or without geographic coordinates), and different clustering algorithms.
We define three criteria for analyzing clustering stability and robustness and conduct a thorough empirical evaluation of the suggested configurations using the data published by ENTSO-E.
Following ACER's methodology, we implement K-Means and Spectral Clustering and evaluate the clusterings computed with these methods.
We want to understand if the clustering algorithms compute the same clusters when nodal prices of different time periods are considered. How different are clusters resulting from different clustering algorithms? Also, how similar are the clusters of a particular configuration with respect to prices, and how large are the price standard deviations within clusters?  

We find that the configurations proposed by ACER are not stable across time and the algorithm used. Moreover, the configurations do not reduce the price standard deviations within zones considerably, and the average prices between zones are similar. If we recompute the clusters and take only prices into account\footnote{This means that we do not include constraints in the clustering computation such that we obtain balanced clusters in terms of either the number of nodes or geographical coverage.}, we can reduce the price standard deviation within clusters further. However, the resulting clusters are not of equal size and are not geographically coherent. Importantly, neither the configurations proposed by ACER nor the clusters that we computed are stable across time. Depending on the time frame taken into account, different clusterings would result. 

The data from the LMP study of the BZR can be considered the most comprehensive data set based on up-to-date information from European TSOs. Based on this data set, splitting Germany into 2, 3 or 4 zones does not lead to large price differences between zones and does not reflect long-term price patterns. Our study analyzes the stability and robustness of the alternative configurations with respect to prices, while other factors, such as network security and the impact on energy transition \citep{acer-methodology}, are relevant to evaluate the positive and negative effects of a bidding zone split. 

There are different reasons for or against a split of the German price zone. The method adopted in the BZR is based on prices from an LMP study and clustering, and it aims for robust and stable zones. This paper does not make an argument for or against a zonal split. However, it shows that finding geographically large and stable price zones might not be possible. 
For the question of whether to split the German price zone and how these price zones should be shaped, the clustering of LMPs allows for many answers.

This paper is structured as follows. In Section \ref{sec:related-literature} we discuss the literature related to our work, while in Section \ref{sec:methodology} we describe the algorithms we implement to compute clusterings and introduce the metrics considered to evaluate stability. 
The data we use in our experiments is discussed in Section \ref{sec:data}, while the experimental results are presented in Section \ref{sec:results}. Finally, we summarize our main findings and their policy implications in Section \ref{sec:conclusion}.

\section{Related Literature} \label{sec:related-literature}
This paper complements existing literature computing bidding zone configurations. In particular, we draw on a unique data set with LMP prices that is provided by ENTSO-E. \cite{bovo2019} identifies the main network indicators that should be considered for defining suitable bidding zone configurations.
LMPs and power transfer distribution factors (PTDFs) represent such indicators. 
While some papers focus on identifying configurations by clustering PTDFs (e.g., \citep{kumar2004} and \citep{klos2014}), most works use LMPs as clustering features \citep{stoft1997}. 
\cite{burstedde2012} computes LMPs for the scenario years 2015 and 2020 considering a simplified 72-node representation of the European grid, and evaluates the shapes and sizes of the zones obtained by clustering these LMPs using hierarchical clustering. 
\cite{breuer2013} implement a genetic algorithm and apply it to a dataset consisting of 380 kV nodes considering a projected system for 2016 and 2018.
Based on six scenarios, \cite{weber2018consistent} compute robust price zone configurations for Central Western Europe by employing hierarchical clustering based on the LMPs of approximately 2,200 nodes and assigning weights to nodes depending on their demand- and supply situation.
These configurations are robust in the sense that LMPs from multiple scenarios are integrated into their computation.  
Considering a network model consisting of around 2,700 nodes, \cite{brouhard2023clustering} design nine zone delineations for Continental Europe by implementing K-Means and hierarchical clustering.
The authors calculate LMPs for multiple, independent single-period optimization problems (i.e., optimal power flow with direct current approximation), and use geographic coordinates and these prices as clustering features.

\cite{sarfati2015five} introduce five indicators to assess the impact of bidding zone configurations in zonally-priced electricity markets. The indicators include commercial exchange and welfare evolution, as well as price convergence and divergence, and were analyzed on the Nordic 32-bus example system.
Using a five-node example with a single congested line, \cite{bjorndal2001} show that defining zones based on nodal prices is not guaranteed to result in a zonal configuration with maximum welfare.
The authors formulate a mixed-integer nonlinear program that, for a given number of zones, computes a zonal configuration with maximum welfare.

In terms of the German-Luxembourgish bidding zone, \cite{zinke2023two} identifies north-south splits using hierarchical clustering based on the LMPs associated to approximately 500 nodes that are computed within a linear market and grid model for four scenario years. 
The author considers uncertain factors such as short-term weather patterns and long-term system changes to evaluate the potential decrease in redispatch costs resulting from such splits.
Moreover, \cite{ambrosius2020endogenous} consider a scenario for the year 2035 and determine welfare-maximizing price zones and available transfer capacities for a simplified representation of the German market area consisting of 28 nodes. Such zones are identified based on a mixed-integer nonlinear model that incorporates a graph partitioning problem on the first level \citep{grimm2019optimal, kleinert2019}. The prior literature used different data sources for these analyses.

\section{Methodology} \label{sec:methodology}
The process of computing price zones can be reduced to solving a clustering problem. Formally, given a set of $N$ $D$-dimensional data points $X = \{x_1, x_2,..., x_N\}$ with $x_i \in \mathbb{R}^D$ and a number of clusters $k$, the objective of the clustering problem is to find for each point $x_i$ the corresponding assignment $z_i \in \{1,..., k\}$ to one of the $k$ clusters such that (1) points within the same cluster are similar to each other and (2) points between different clusters are dissimilar from each other. 
A clustering is a solution to this problem and is denoted by $\mathcal{C} = \{C_1, \dots, C_k\}$, where $C_i = \{x_j: z_j = i\}$ is a cluster.
The centroid of a cluster $C_i$ is defined as the mean of the data points assigned to $C_i$ and is depicted by $c_i$. 
In our setting, the data points are the nodes in the transmission network and the dimensions represent the features we use to compute clusters (see Section \ref{sec:feature-selection}).

\subsection{Clustering Algorithms} \label{sec:algorithms}
Following ACER's approach to identify alternative price zone configurations, we consider two clustering algorithms in this paper. 
ACER used three clustering approaches \citep{acer-clustering-algo}, i.e., \textcolor{black}{constrained} K-Means, Spectral Clustering with constrained K-Means and Spectral Clustering with CDIRC, but the implementation details are not publicly available. 
Therefore, it is impossible to reproduce the exact algorithmic configuration ACER implemented. Instead, we consider the same family of clustering methods to evaluate alternative price zone configurations. In what follows, we briefly describe the clustering algorithms we implement.

\subsubsection{\textcolor{black}{Constrained} K-Means} \label{sec:k-means}
K-Means \citep{kmeans-lloyd1982} is a simple algorithm that clusters data points by minimizing within-cluster variances, i.e., squared Euclidean distances. The method consists of three main steps. In the first step, the centroids are initialized randomly. In the second step, each data point is assigned to its nearest centroid with respect to the Euclidean distance, specifically to its corresponding cluster. The third step consists of updating the centroids, given the new points assigned to the clusters.
The last two steps are repeated until the difference between the old and the new centroids is below a threshold. 
Importantly, the final clusters might differ depending on the initialization of centroids.

A constrained version of K-Means was considered in the BZR analysis \citep{acer-clustering-algo}, in which the centroids are initialized such that they are spread out spatially. We note that this version of K-Means is known as K-Means++ \citep{k-means++}.
We implement the constrained version of K-Means by \cite{bennett2000constrained} and incorporate constraints for the minimum size of points included in each cluster. 
We update the clusters until the squared Euclidean distance between each old and new centroid is smaller than 0.1.
Moreover, we standardize the data such that all features have a mean of 0 and a standard deviation of 1.

\subsubsection{Spectral Clustering \textcolor{black}{with Constrained K-Means}} \label{sec:sc}
Spectral clustering is a technique used for clustering network data. Given a graph $G = (V, E)$, the method clusters the set of vertices $V$ based on a low-dimensional spectral embedding of $G$ and consists of three main steps. 
First, the Laplacian matrix $L$ is constructed. Then, the first $r$ eigenvectors (i.e., the eigenvectors corresponding to the $r$ smallest eigenvalues of $L$) are computed. In the last step, a matrix $\tilde{X} \in \mathbb{R}^{n \times r}$ is considered, whose columns are the selected eigenvectors and the $i$-th row is an embedding of the $i$-th node in $G$.
The graph nodes are then clustered by applying, e.g., K-Means on the node embeddings. We refer to \cite{aggarwal.2021.spectral} for a comprehensive overview of Spectral Clustering.

For our setting, we define a complete graph $G$ in which $V$ is the set of nodes considered in the BZR analysis. We set the weight $w_e$ of an edge $e = (u, v) \in E$ as the similarity between nodes $u$ and $v$ and compute it using the radial basis function kernel. Thus, with $x_u, x_v$ denoting the features corresponding to nodes $u$ and $v$, respectively, we define $w_e = \exp(-\frac{\lVert x_u - x_v \rVert^2}{2\sigma^2})$, where $\sigma$ is a parameter that controlls the width of node-neighborhoods. We normalize the edge weights such that $w_e \in [0,1]$ for all $e \in E$ and set $\sigma = 1$.
To cluster the nodes using $\tilde{X} \in \mathbb{R}^{n \times r}$, we apply the constrained version of K-Means mentioned in Section \ref{sec:k-means}. Moreover, we set $r = k$ and thus consider the first $k$ eigenvectors as clustering features.

\subsection{Stability Evaluation} \label{sec:stability-evaluation}
In what follows, we discuss three criteria that we consider to evaluate the stability of pricing zones.
Given a period $t$ and a node $x$, we denote by $x^t$ the vector containing the time series of prices corresponding to $x$ in period $t$. We define $\overline{x}^t$ as the average price of node $x$ in period $t$, which is simply the average over all values in $x^t$. 
We also denote the distance between two points $x$ and $y$ in period $t$ by $d^t(x, y)$ and define it as $d^t(x, y) = \lvert \overline{x}^t - \overline{y}^t \rvert$.
Whenever the period we are referring to is evident, we simplify notation by omitting the reference to $t$ in $d^t(x, y)$.

\subsubsection{Intra- and Inter-Cluster Similarity} \label{sec:intra-inter-similarity}
In this section, we assume that a clustering $\mathcal{C} = \{C_1, \dots, C_k\}$ is given for a period $t$. 
We consider $\mathcal{C}$ to be stable if it is comprised of clusters having points that are similar with respect to their prices. In other words, $\mathcal{C}$ has large intra-cluster similarity.
To analyze intra-cluster similarity, we consider the \textit{price standard deviation} of each cluster $C_i$ and denote it by $\sigma_i$. We compute $\sigma_i$ based on the vectors $x^t$ corresponding to the nodes assigned to $C_i$. 
The \textit{average} price standard deviation $\sigma (\mathcal{C})$ of a clustering $\mathcal{C}$ is the average of all cluster price standard deviations $\sigma_i, C_i \in \mathcal{C}$. 
Whenever the clustering we are referring to is clear, we simplify notation by dropping the reference to $\mathcal{C}$.
A low average price standard deviation indicates large intra-cluster similarity.

Furthermore, a stable clustering is characterized by low inter-cluster similarity or, equivalently, large inter-cluster dissimilarity. 
With $\mu_i$ denoting the mean price of a cluster $C_i$\footnote{Note that, similar to $\sigma_i$, $\mu_i$ is computed based on the vectors containing the time series of prices corresponding to the nodes assigned to $C_i$.}, we define the following distance measures to assess the dissimilarity between two clusters $C_i, C_j \in \mathcal{C}$: 

\noindent
The \textit{average distance between cluster mean prices} quantifies the distinctiveness of clusters based on mean prices: 
$$d_\mu (C_i, C_j) = d(\mu_i, \mu_j).$$
The \textit{nearest neighbor cluster distance} reflects how well-separated the clusters are: 
$$d_{\min} (C_i, C_j) = \min_{x \in C_i, y \in C_j} d(x, y).$$ 
The \textit{average distance between clusters} considers all pairs of points to measure the dissimilarity between $C_i$ and $C_j$:
$$d_{\text{avg}} (C_i, C_j) = \frac{1}{|C_i| \cdot |C_j|} \sum\limits_{x \in C_i} \sum\limits_{y \in C_j} d(x, y).$$
The following quantity is used to evaluate the \textit{global} inter-cluster dissimilarity of $\mathcal{C}$: 
$$\Delta_{q} (\mathcal{C}) = \frac{1}{\binom{K}{2}} \sum\limits_{\substack{C_i, C_j \in \mathcal{C} \\ i < j}} d_q (C_i, C_j),$$
where $q$ is one of \{$\mu$, min, avg\}.

We also consider the \textit{Davies-Bouldin index} to assess both the intra-cluster similarity and inter-cluster dissimilarity of $\mathcal{C}$. 
With $s_i$ denoting the diameter\footnote{The diameter of a cluster is the maximum distance between any two points assigned to this cluster.} of cluster $C_i$ and $d_{ij}$ the distance between $c_i$ and $c_j$, the Davies-Bouldin Index [\citenum{davies-index}] is defined as:
$$DB = \frac{1}{k} \sum\limits_{i=1}^{K} \max_{i \neq j} R_{ij},$$
were $R_{ij} = \frac{s_i + s_j}{d_{ij}}$.
Intuitively, $DB$ quantifies the average similarity of every cluster with its most similar cluster. Here, similarity is defined as the ratio of within-cluster distances to between-cluster distances. 
A low $DB$ index indicates a good clustering, and the minimum value is 0. 
Clusters that are well-separated and less dispersed are rewarded by the $DB$ index.

\subsubsection{Temporal Stability} \label{sec:temportal-stability}
While geographic coordinates are static across time, prices are time-sensitive. Therefore, we are interested to see if pricing zones are stable over time. To decide this, we compare two clusterings $\mathcal{C}_1$ and $\mathcal{C}_2$ that are computed for two different time periods with the same algorithm and considering the same features. For this comparison, we use the \textit{Rand index}, which measures the similarity between $\mathcal{C}_1, \mathcal{C}_2$ considering all pairs of data points and counting the pairs that are assigned to the same or different clusters in both $\mathcal{C}_1$ and $\mathcal{C}_2$:
$$RI(\mathcal{C}_1, \mathcal{C}_2) = \frac{\sum\limits_{1 \leq l < r \leq N} \gamma(x_l, x_r)}{\binom{N}{2}},$$
where 
\begin{equation*}
	\gamma(x_l, x_s) =
	  \begin{cases}
		1 & \text{if $x_l$ and $x_r$ are in the same cluster in $\mathcal{C}_1$ and in $\mathcal{C}_2$}\\
		1 & \text{if $x_l$ and $x_r$ are in different clusters in $\mathcal{C}_1$ and in $\mathcal{C}_2$}\\
		0 & \text{otherwise.}
	  \end{cases}       
\end{equation*}
Rand indices are between 0 and 1, and 1 indicates a perfect match, that is, the same clustering is obtained in both time periods.

\subsubsection{Spatial Coherence}
We evaluate clustering stability also with respect to whether the clusters are geographically separated.
Geographically separated or spatially cohesive clusters can be identified on the map as groups of points (nodes) that are geographically close.  

We assess whether clusters are spatially cohesive considering spatial autocorrelation. Thus, for a clustering $\mathcal{C}$, a positive spatial autocorrelation indicates that $\mathcal{C}$ has spatially cohesive clusters since neighboring locations are more likely to share similar prices and belong to the same cluster.
To measure spatial autocorrelation, we use \textit{Global Moran's I} \citep{moran1950}, which is defined as:
$$I = \frac{N}{W} \frac{\sum\limits_{x \in X}\sum\limits_{y \in X}w_{xy} (c(x) - \overline{c})(c(y) - \overline{c})}{\sum\limits_{x \in X} (c(x) - \overline{c})^2},$$
where $c(x)$ is the centroid of the cluster to which point $x$ is assigned, $w_{xy}$ denotes the spatial distance between points $x, y$ and $W = \sum\limits_{x \in X}\sum\limits_{y \in X}w_{xy}$. Also, $\overline{c}$ represents the average of all centroids.
$I$ values are between -1 and +1, where +1 defines perfect clustering of similar values, 0 indicates no correlation and -1 suggests perfect clustering of dissimilar values.

\section{Data} \label{sec:data}
The data we use in our experiments was published as part of the locational pricing study conducted by ENTSO-E\footnote{The datasets can be accessed at \url{https://www.entsoe.eu/network_codes/bzr/} (January 2024).} and includes locational (i.e., nodal) prices.
In this study, 24 weeks spanned across 3 \textcolor{black}{scenario} years were analyzed: 1989 (weeks 04, 10, 11, 17, 20, 31, 40, 52), 1995 (weeks  02, 12, 16, 21, 27, 36, 38, 49) and 2009 (weeks 04, 08, 11, 15, 16, 21, 31, 48).
For each day in these weeks, nodal prices for every 2-hour interval are available. The file containing the prices for the German-Luxembourgish bidding zone includes 6,090 unique node IDs and 14,170,464 nodal prices. Moreover, for 1,893,024 ($\approx13\%$) nodal prices, which correspond to 939 distinct nodes, the node ID is ``nan". Consequently, these prices cannot be considered in our experiments, as insufficient data is provided for identifying the nodes to which these prices correspond.

Note that ENTSO-E did not publish the geographic coordinates of the nodes analyzed in the locational pricing study. However, these are essential for identifying and analyzing price zones. Therefore, we implement the following approach to obtain the geographic location of nodes. 
We consider the XML files describing the German grid model that we obtained from the Bidding Zone Review website \citep{bz-website}. The grid is composed of elements such as substations, voltage levels and topological nodes, which form a hierarchy as follows: substations consist of one or multiple voltage levels and voltage levels consist of one or multiple topological nodes.
Altogether, the German grid encompasses 834 substations, 1,697 voltage levels and 2,898 topological nodes. 
The name of each substation is included in the XML files and based on this we identified the geographic location of each substation using OpenStreetMap and (JAO) Static Grid Model\footnote{\url{https://www.jao.eu/static-grid-model}}.
Since a topological node is mapped to a specific substation, in this process we also obtain the geographic coordinates of the topological nodes.

Next, we identify the prices corresponding to the topological nodes. We do this based on the IDs of these nodes and the IDs of the 6,090 nodes included in the file containing LMPs. 
For 2,200 out of 2,898 topological nodes we could identify their IDs in this file. These nodes belong to 714 substations across Germany. Looking at the different LMP prices computed by ENTSO-E ordered by substation, it turns out that prices for different topological nodes associated with a substation are the same or very close. So, with the 714 substations covered in our study, we could reconstruct most of the German substations, their prices and locations across Germany.

\subsection{Clustering Features \& Periods} \label{sec:feature-selection}
We apply K-Means and Spectral Clustering for two feature configurations. In the first configuration we use time series of nodal prices as clustering features, that is, a feature represents the nodal price of a specific point in time.
In the second configuration, we additionally consider latitude and longitude.
We compute one clustering for every \textcolor{black}{scenario} year analyzed in the bidding zone review, i.e., 1989, 1995 and 2009. For each year, 672 prices are reported per node ($12 \text{ hours} \times 7 \text{ days} \times 8 \text{ weeks}$), which implies that 672 and 674 features are considered for the first and second configuration, respectively. Note that the first configuration was also used by ACER \citep{acer-approach-clustering}.

\section{Results} \label{sec:results}
In this section, we first analyze the price zone configurations proposed by ACER. Then, we evaluate the optimal price zone configurations computed with \textcolor{black}{our implementation of} K-Means and Spectral Clustering for the \textcolor{black}{scenario} years 1989, 1995 and 2009. Considering the metrics defined in Section \ref{sec:stability-evaluation}, we assess the stability of the resulting clusterings and compare them to those proposed by ACER.
\textcolor{black}{In what follows, whenever we mention K-Means and Spectral Clustering as clustering methods, we refer to our implementation described in Section \ref{sec:algorithms}.}

\subsection{Analysis of the price zone configurations proposed by ACER}
\textcolor{black}{The configurations proposed by ACER for Germany are specified in Annex I of \textit{ACER's Decision on the alternative bidding zone configurations to be considered in the bidding zone review process} \citep{acer-list-all-proposed-bzs}. For brevity reasons, we refer to the two-zone configurations computed using K-Means and Spectral Clustering as DE2 (k-means) and DE2 (spectral), respectively. Similarly, the configurations with three and four zones are denoted as DE3 (spectral) and DE4 (spectral).}
Appendix \ref{app:approx-bz-acer} contains approximations of these configurations given the nodes we consider in our experiments and based on the maps provided by ACER \citep{acer-list-all-proposed-bzs}. Since it is difficult to manually obtain an accurate geographic separation of the zones within the proposed configurations, the configurations from Appendix \ref{app:approx-bz-acer} are just approximations.  
In what follows, we analyze these configurations.

\begin{figure}[!htp]
	\centering
	\includegraphics[width=.3\textwidth]{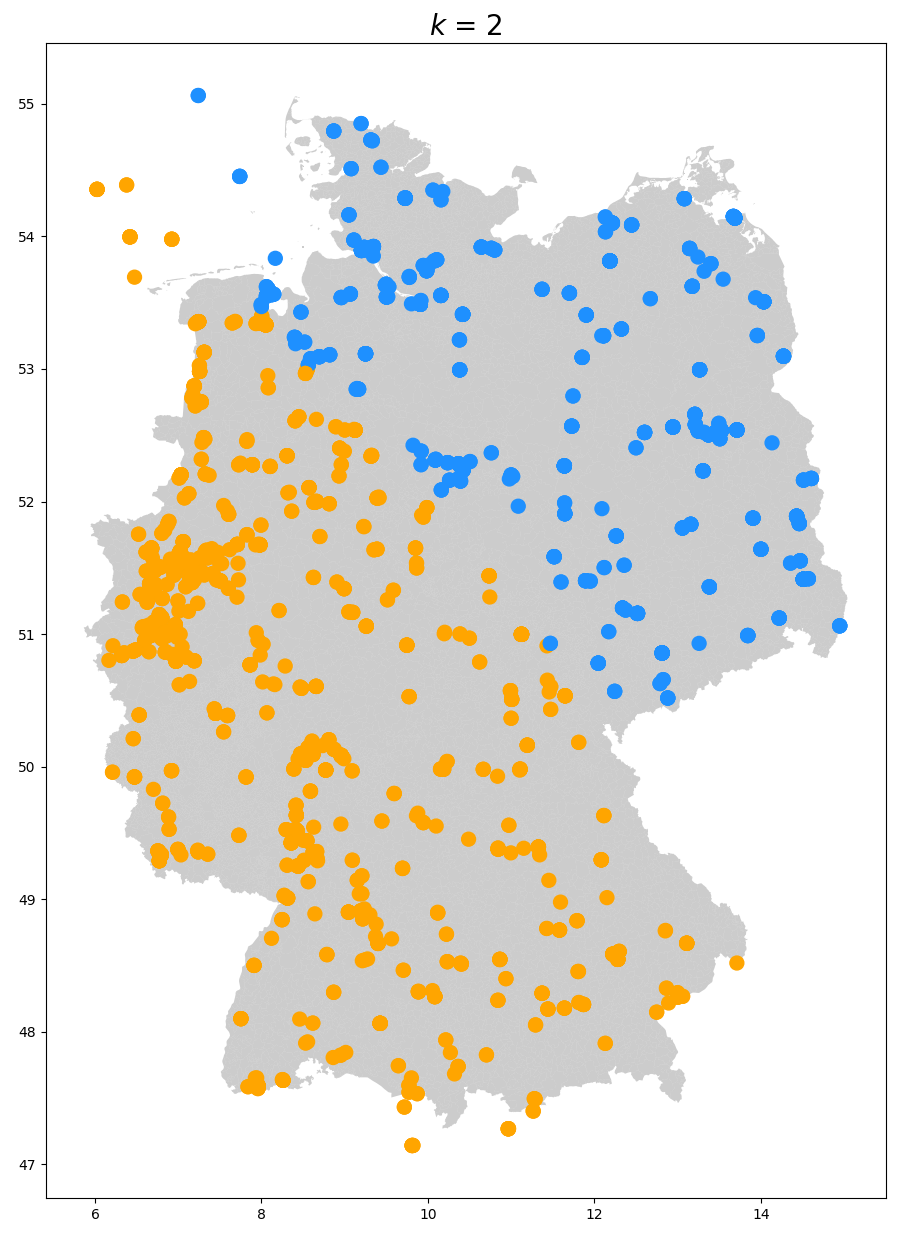}
	\includegraphics[width=.3\textwidth]{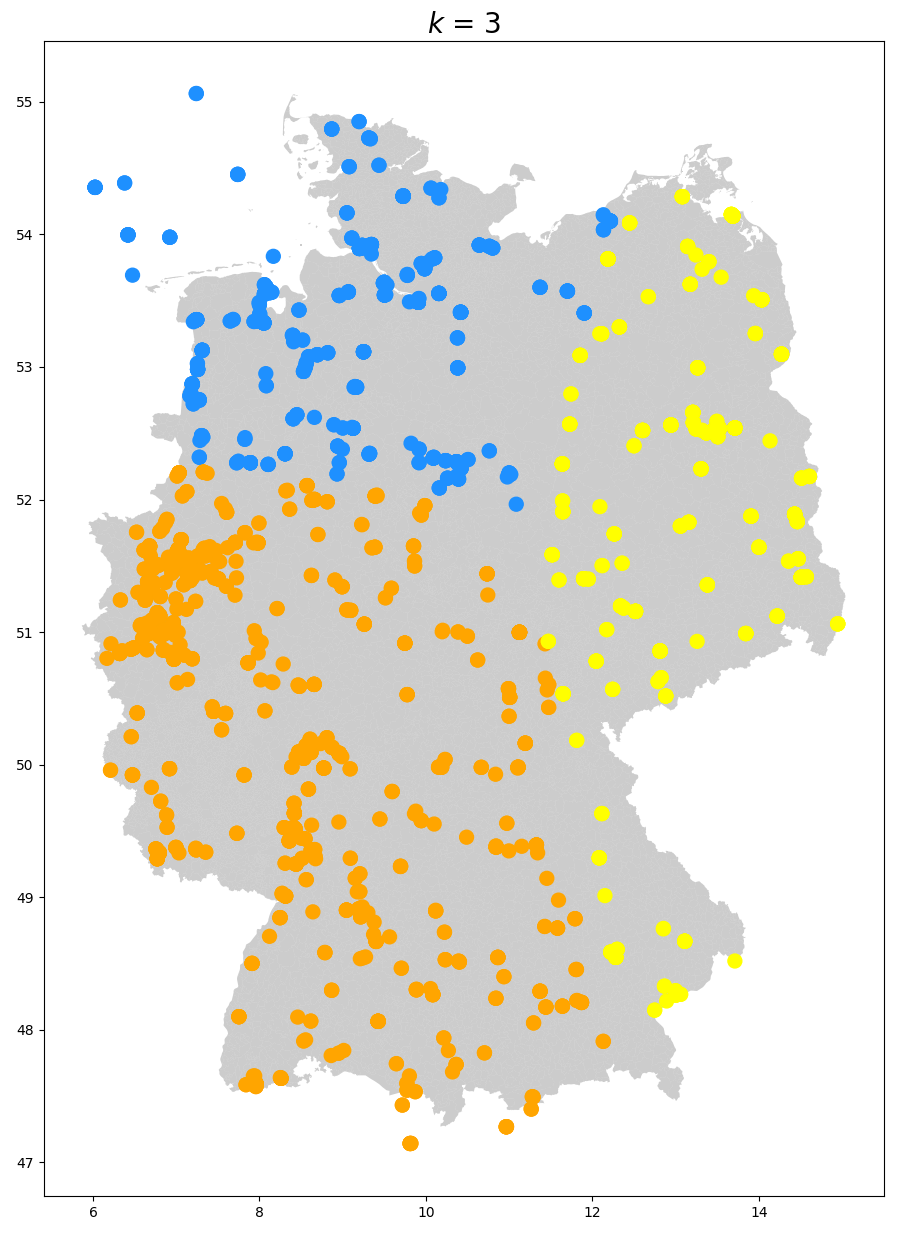}
	\includegraphics[width=.3\textwidth]{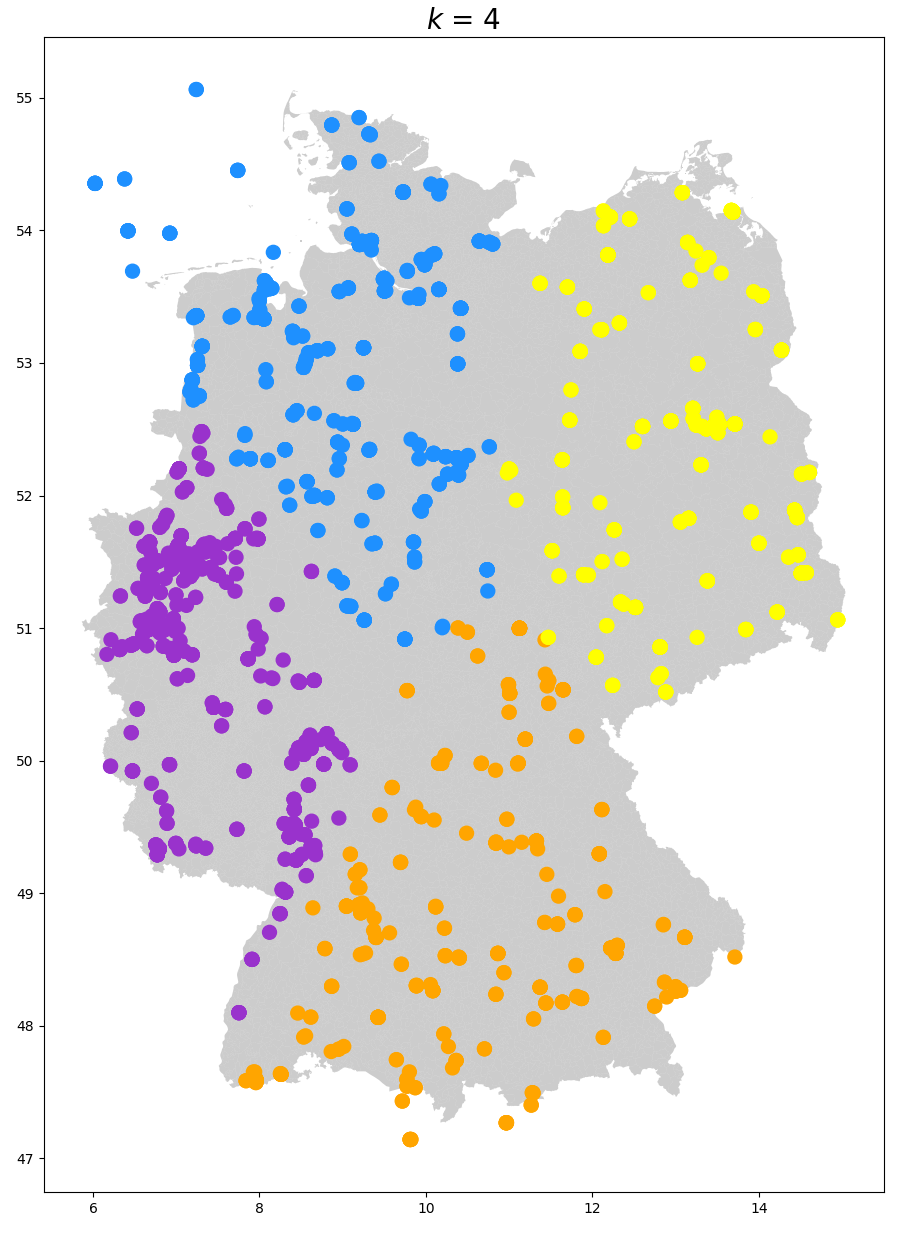}
	\caption{Price zones computed with K-Means for $k \in \{2, 3, 4\}$ based on latitude and longitude}
	\label{fig:bz_lat_long_k_means}
\end{figure}

First, we want to understand the extent to which the geographic coordinates of the nodes contribute to the determination of price zones.
To do this, we apply K-Means with latitude and longitude as features\footnote{We also imposed a threshold related to the minimum number of nodes to be included in a zone.} to cluster the nodes, but do not consider prices. 
The computed clusters are shown in Figure \ref{fig:bz_lat_long_k_means}.

\begin{result}
	Clustering nodes based solely on latitude and longitude leads to price zones that closely resemble the configurations proposed by ACER.   
\end{result}

We notice that DE2 (k-means), DE3 (spectral) and DE4 (spectral) are similar to the clusterings computed by K-Means. In fact, the configurations proposed by ACER seem to be adjustments of the zones illustrated in Figure \ref{fig:bz_lat_long_k_means} such that their borders do not cross any federal state, with few exceptions at the borders between North Rhein-Westphalia \& Lower Saxony (for DE2 (spectral)) and Hesse \& North Rhein-Westphalia \& Lower Saxony (for DE3 (spectral) and DE4 (spectral)). Further below, we discuss clusterings that take prices into account and compare them to those of ACER. 

Before this, we analyze the nodal prices for \textcolor{black}{scenario years} 1989, 1995 and 2009 in the configurations proposed by ACER.  
For each year and week, we compute the cluster price standard deviations and average prices corresponding to configurations DE2 (k-means), DE2 (spectral), DE3 (spectral) and DE4 (spectral), respectively. 
The purpose of this analysis is to see what benefit multiple price zones have on the reduction of the price standard deviation inside zones. Furthermore, we study whether the clusters are well-defined, i.e., whether, for a given configuration, the cluster average prices are different.

The evolution of cluster price standard deviations across the weeks in \textcolor{black}{the scenario years} 1989, 1995 and 2009 is illustrated in Appendix \ref{app:cluster-variances}. We also observe the price standard deviations corresponding to a single price zone or cluster (Figure \ref{fig:1-zone-price-var}).
Clearly, the average cluster price standard deviations decrease as we consider configurations with 2, 3 or 4 clusters. This is an implicit objective of clustering nodes based on prices. 
However, this reduction is not substantial and varies between 3\% to 9\% depending on the configuration. For instance, in 2009, the percentage decreases in average price standard deviations compared to a single zone range from 1.92 to 5.69. 
Moreover, the decrease in average cluster price standard deviation is not monotonic with respect to the number of clusters. That is, a larger number of clusters does not necessarily imply smaller average cluster price standard deviations.
For example, for 2009 and configuration DE2 (k-means), the average price standard deviation is 10.85, while DE3 (spectral) leads to an average price standard deviation of around 11.26.
An explanation for this behavior is that the average cluster prices are heavily influenced by outliers, which are nodes having very large or low prices compared to the majority of nodes.  
For smaller clusters, that is, for configurations with a larger number of clusters, the influence of outliers is stronger. 
Moreover, if we compare the configuration with one price zone to any configuration with multiple zones (clusters), we notice that a reduction in average price standard deviation does not necessarily imply that \textit{all} cluster standard deviations are lower. 

Next, we analyze how different the average cluster prices are for each configuration. These values are shown in Appendix \ref{app:cluster-average-prices}. 

\begin{result}
	The difference in the average prices of clusters in the proposed configurations is low. For all configurations, the yearly differences between the average cluster prices are less than 6 EUR/MWh and 4.16 EUR/MWh on average. 
\end{result}

The differences between the average prices in Germany and the average prices in the neighboring countries are generally larger than 6 EUR/MWh.\footnote{\url{https://www.energy-charts.info/charts/price_average_map/chart.htm?l=en&c=DE&year=2018&interval=year}}
The German-Austrian price zone was split in 2018. The average electricity spot market price in 2018 was 41.74 EUR/MWh in Germany and 59.92 EUR/MWh in Austria, which implies a price difference of 18.18 EUR/MWh between these zones.
Our analysis indicates that the configurations proposed by ACER do not lead to well-defined clusters in terms of average prices. As the tables in Appendix \ref{app:cluster-average-prices} showcase, the cluster average prices are generally similar, regardless of the configuration, week or year.

Let us now compare the configurations proposed by ACER to the clusterings computed with our implementation of K-Means and Spectral Clustering. Given the nodal prices for \textcolor{black}{the scenario years} 1989, 1995 and 2009, respectively, we apply both clustering algorithms to identify clusterings (i.e., price zones) for each year.
The goal of this comparison is to understand whether the proposed configurations by ACER are stable over time.
We say that the ACER configurations are \textit{stable} if they are similar to the optimal clusterings identified for different years.

As described in Section \ref{sec:feature-selection}, we compute clusterings for two scenarios. The first scenario uses, as clustering features, the time series of nodal prices, while the second scenario additionally considers the location coordinates. 
Table \ref{tab:compare-official-bz-rand} shows Rand indices measuring the similarity between the clusterings proposed by ACER and the clusterings we computed with each algorithm for \textcolor{black}{the scenario years} 1989, 1995 and 2009. 
For example, the first Rand index (0.5, column 3) indicates the similarity between (1) the solution computed with our implementation of K-Means using the time series of nodal prices corresponding to the scenario year 1989 and (2) configuration DE2 (k-means) proposed by ACER.

\begin{table}[!htp]
	\centering
	\begin{tabular}{c|c|c|c|c||c|c|c}
		& & \multicolumn{3}{c||}{Without loc.} & \multicolumn{3}{c}{With loc.} \\ 
		& & 1989 & 1995 & 2009 & 1989 & 1995 & 2009 \\
        \hline
        \multirow{4}{*}{K-Means} & ACER DE2 (k-means) & 0.5 & 0.5 & 0.51 & 0.5 & 0.5 & 0.51 \\
        & ACER DE2 (spectral) & 0.5 & 0.5 & 0.5 & 0.5 & 0.5 & 0.5 \\
        & ACER DE3 (spectral) & 0.51 & 0.53 & 0.52 & 0.52 & 0.52 & 0.52 \\
        & ACER DE4 (spectral) & 0.6 & 0.5 & 0.56 & 0.6 & 0.58 & 0.57 \\
        \hline
        \multirow{4}{*}{Spectral} & ACER DE2 (k-means) & 0.51 & 0.51 & 0.51 & 0.51 & 0.51 & 0.51 \\
        & ACER DE2 (spectral) & 0.5 & 0.5 & 0.5 & 0.5 & 0.5 & 0.5 \\
        & ACER DE3 (spectral) & 0.52 & 0.44 & 0.52 & 0.52 & 0.44 & 0.52 \\
        & ACER DE4 (spectral) & 0.55 & 0.45 & 0.57 & 0.56 & 0.46 & 0.56 \\
	\end{tabular}
	\caption{Comparison between the alternative price configurations computed with our implementation of K-Means and Spectral Clustering and the price zones proposed by ACER. The table shows Rand indices, which are in $[0, 1]$, 1 indicating perfect similarity. We computed clusterings using K-Means and Spectral Clustering using: (1) time series of nodal prices (columns ``Without loc.'') and (2) time series of nodal prices and the location coordinates of nodes (column ``With loc.'').}
	\label{tab:compare-official-bz-rand}
\end{table}

\begin{result}
	The price zone configurations proposed by ACER are not stable over time. Given the nodal prices corresponding to \textcolor{black}{the scenario years} 1989, 1995 and 2009, respectively, K-Means and Spectral Clustering compute price zones that are considerably different from the proposed configurations.
\end{result}

We recall from Section \ref{sec:stability-evaluation} that the Rand index measures the similarity between two clusterings by considering all pairs of nodes and counting the pairs that are assigned to the same or different clusters in both clusterings.
As Table \ref{tab:compare-official-bz-rand} indicates, Rand indices vary considerably when different years are analyzed. For instance, for DE4 (spectral) and the clustering computed with Spectral Clustering with price and location features we get $RI=0.56$ for 1989 \textcolor{black}{yr scenario}, $RI=0.46$ for 1995 \textcolor{black}{yr scenario} and $RI=0.56$ for 2009 \textcolor{black}{yr scenario}.
Thus, our analysis suggests that the configurations proposed by ACER are not stable over time, regardless of the number of price zones (see also Appendix \ref{app:rand-visualization}).

\color{black}
There are two main reasons for temporally unstable configurations.
The first reason is that nodal prices vary both temporally and spatially due to various factors, such as supply and demand fluctuations, fuel prices and weather conditions. 
The second reason is related to the non-robustness of the clustering algorithms that we consider.
K-Means is sensitive to the initial placement of centroids and outliers, which can significantly affect the resulting clusters.
Spectral Clustering also exhibits sensitivity to the choice of parameters such as the implemented similarity metric.
\color{black}

\subsection{Analysis of the optimal price zone configurations for \textcolor{black}{the scenario years} 1989, 1995 and 2009}
In what follows, we analyze alternative price zones computed with K-Means and Spectral Clustering for the \textcolor{black}{scenario} years 1989, 1995 and 2009. Previously we showed that the configurations proposed by ACER are not stable over these years. Now we are interested to see whether other price zone configurations exist that are more stable with regards to the criteria defined in Section \ref{sec:stability-evaluation}. 
We report the results of K-Means and Spectral Clustering as described in Section \ref{sec:algorithms}.
We evaluate the clusterings computed for each year using (1) time series of nodal prices as features, and (2) time series of nodal prices and the location coordinates of nodes.

Based on the results provided in Appendix \ref{app:intra-inter}, we first evaluate the intra-cluster similarity and the inter-cluster dissimilarity of the clusterings computed for all three years. For each year, we want to understand whether the clusters are well-defined. Well-defined clusters are characterized by low price standard deviation within clusters (or, large intra-cluster similarity) and different cluster average prices (or, large inter-cluster dissimilarity). 

\begin{result}
	The clusterings computed with K-Means for \textcolor{black}{the scenario years} 1989, 1995 and 2009 based only on prices exhibit lower average price standard deviations compared to the configurations proposed by ACER.  
\end{result}

Tables \ref{tab:2-zones}, \ref{tab:3-zones} and \ref{tab:4-zones} show the average price standard deviations for different configurations with 1, 2, 3 or 4 price zones or clusters. The configurations refer to (1) a single price zone (column `Single price zone'), (2) the configurations proposed by ACER (columns ACER DE2 (k-means), DE2 (spectral), DE3 (spectral), DE4 (spectral)) and (3) the optimal configurations computed with K-Means and Spectral Clustering based on nodal prices (columns `K-Means' and `Spectral Clustering').  

\begin{table}[!htp]
    \centering
    \begin{tabular}{c|c|c|c|c|c}
        & Single price zone & ACER DE2 (k-means) & ACER DE2 (spectral) & K-Means & Spectral  \\
        \hline
        1989 & 14.54 & 13.09 & 13.98 & 13.40 & 13.39 \\ 
        1995 & 14.31 & 12.93 & 13.80 & 11.16 & 15.16 \\
        2009 & 11.55 & 10.85 & 11.33 & 10.80 & 11.61 \\
    \end{tabular}
    \caption{Average price standard deviations of configurations with 1 or 2 price zones}
    \label{tab:2-zones}

	\vspace{0.5cm}

    \centering
    \begin{tabular}{c|c|c|c|c}
        & Single price zone & ACER DE3 (spectral) & K-Means & Spectral  \\
        \hline
        1989 & 14.54 & 13.86 & 12.11 & 14.81 \\ 
        1995 & 14.31 & 13.65 & 12.47 & 15.56 \\
        2009 & 11.55 & 11.26 & 10.06 & 11.11 \\
    \end{tabular}
    \caption{Average price standard deviations of configurations with 1 or 3 price zones}
    \label{tab:3-zones}

	\vspace{0.5cm}

    \centering
    \begin{tabular}{c|c|c|c|c}
        & Single price zone & ACER DE4 (spectral) & K-Means & Spectral  \\
        \hline
        1989 & 14.54 & 13.14 & 12.86 & 12.23 \\ 
        1995 & 14.31 & 13.02 & 11.12 & 15.09 \\
        2009 & 11.55 & 10.87 & 10.05 & 10.73 \\
    \end{tabular}
    \caption{Average price standard deviations of configurations with 1 or 4 price zones}
    \label{tab:4-zones}
\end{table}

We observe that the average price standard deviations $\sigma$ differ depending on the clustering algorithm. For instance, for $k=3$, K-Means (Spectral Clustering) leads to $\sigma=12.11 (14.81)$ for 1989 \textcolor{black}{yr scenario}, $\sigma=12.47 (15.56)$ for 1995 \textcolor{black}{yr scenario} and $\sigma=10.06 (11.11)$ for 2009 \textcolor{black}{yr scenario}.
However, regardless of the year and the number of clusters, the clusterings we compute with K-Means based on nodal prices lead to average price standard deviations that are generally lower compared to the configurations proposed by ACER. 
Recall that the average price standard deviations for the configurations proposed by ACER are up to 13.98. 
Nevertheless, the clusters we compute with K-Means and Spectral Clustering are not necessarily balanced, that is, the clusters may not contain a similar number of nodes. This fact can be observed in the plots provided in Appendix \ref{app:rand-visualization}.
Given this observation and the fact that the configurations proposed by ACER appear to be balanced (see Appendix \ref{app:approx-bz-acer}), it explains that the price standard deviations inside the clusters we compute with K-Means and Spectral Clustering are lower.

\begin{result}
	The clusterings computed with K-Means and Spectral Clustering for \textcolor{black}{the scenario years} 1989, 1995 and 2009 are characterized by low inter-cluster dissimilarity and are thus not well-defined. Regardless of algorithm or clustering features, the yearly differences between the average cluster prices are less than 6 EUR/MWh and 3.45 EUR/MWh on average. 
\end{result}

To evaluate inter-cluster dissimilarity, we analyze the values of $\Delta_\mu$, $\Delta_{\min}$ and $\Delta_{\text{avg}}$ included in Appendix \ref{app:intra-inter}. We recall from Section \ref{sec:intra-inter-similarity} that $\Delta_\mu$ denotes the global average distance between cluster mean prices, $\Delta_{\text{min}}$ represents the global nearest neighbor cluster distance, while $\Delta_{\text{avg}}$ encodes the global average distance between clusters. For a given period, the distance between two nodes is simply the absolute difference between their average prices in this period.
We notice that, regardless of the year, algorithm or clustering features, $\Delta_\mu$ is smaller than 6 EUR/MWh, which suggests that, in terms of average prices, clusters are rather similar to each other.
Moreover, all values of $\Delta_{\min}$ are smaller than 1, indicating that clusters are not well-separated. This means that there exist nodes that have very similar prices (i.e., their price difference is less than 1) that are assigned to different clusters. 
We also observe that the values of $\Delta_{\text{avg}}$ are in the interval $[2.51, 5.20]$ and become slightly smaller as we consider configurations with multiple clusters. In other words, the average price distance between any pair of two nodes assigned to different clusters is below 6 EUR/MWh. 

Finally, we analyze the Davies-Bouldin indices to measure the average similarity of every cluster with its most similar cluster.
We notice that these indices are quite large and fluctuate depending on the clustering algorithm and clustering features, ranging from 0.45 to 7.75. We recall from Section \ref{sec:intra-inter-similarity} that Davies-Bouldin indices are in $[0, \infty)$, where 0 indicates a good clustering having well-separated and less dispersed clusters.
Interestingly, there is no clear evidence that using location coordinates in the clustering computation leads to smaller indices. 

Given these observations, we conclude that the price zones computed for \textcolor{black}{the scenario years} 1989, 1995 and 2009 with K-Means and Spectral Clustering generally exhibit larger intra-cluster similarity and similar inter-cluster dissimilarity compared to the configurations proposed by ACER.
Large intra-cluster similarity indicates strong cohesion inside clusters and is a consequence of unbalanced clusters relative to the ACER configurations.

Next, we evaluate temporal stability considering the Rand indices included in Table \ref{tab:temp-stability-rand}.

\begin{table}[!htp]
	\centering
	\begin{tabular}{c|c|c|c|c|c|c|c|c|c|c}
		& & \multicolumn{3}{c|}{1989} & \multicolumn{3}{c|}{1995} & \multicolumn{3}{c}{2009} \\
        & & $k=2$ & $k=3$ & $k=4$ & $k=2$ & $k=3$ & $k=4$ & $k=2$ & $k=3$ & $k=4$ \\
        \hline
        \multirow{3}{*}{K-Means} & 1989 & - & - & - & 0.57 & 0.69 & 0.7 & 0.87 & 0.88 & 0.83 \\
        & 1995 & 0.57 & 0.69 & 0.7 & - & - & - & 0.61 & 0.78 & 0.67 \\
        & 2009 & 0.87 & 0.88 & 0.83 & 0.61 & 0.78 & 0.67 & - &  - & - \\
		\hline
		\multirow{3}{*}{Spectral} & 1989 & - & - & - & 0.68 & 0.5 & 0.5 & 0.72 & 0.48 &  0.78\\
		& 1995 & 0.68 & 0.5 & 0.5 & - & - & - & 0.7 & 0.48 & 0.44 \\
		& 2009 & 0.72 & 0.48 & 0.78 & 0.7 & 0.48 & 0.44 & - & - & - \\
	\end{tabular}

	\vspace{0.5cm}
	
	\begin{tabular}{c|c|c|c|c|c|c|c|c|c|c}
		& & \multicolumn{3}{c|}{1989} & \multicolumn{3}{c|}{1995} & \multicolumn{3}{c}{2009} \\
        & & $k=2$ & $k=3$ & $k=4$ & $k=2$ & $k=3$ & $k=4$ & $k=2$ & $k=3$ & $k=4$ \\
        \hline
		\multirow{3}{*}{K-Means} & 1989 & - & - & - & 0.95 & 0.95 & 0.76 & 0.59 & 0.93 & 0.86 \\
		& 1995 & 0.95 & 0.95 & 0.76 & - & - & - & 0.57 & 0.92 & 0.66 \\
		& 2009 & 0.59 & 0.93 & 0.86 & 0.57 & 0.92 & 0.66 & - & - & - \\
		\hline
        \multirow{3}{*}{Spectral} & 1989 & - & - & - & 0.69 & 0.5 & 0.46 & 0.71 & 0.78 & 0.76 \\
        & 1995 & 0.69 & 0.5 & 0.46 & - & - & - & 0.7 & 0.49 & 0.5 \\
        & 2009 & 0.71 & 0.78 & 0.76 & 0.7 & 0.49 & 0.5 & - & - & - \\
	\end{tabular}

	\caption{Temporal stability evaluation. Tables show Rand indices, which are always in $[0, 1]$, 1 indicating perfect similarity. In the first table, the results refer to clusterings computed based on prices. In the second table, the results relate to clusterings computed based on prices and location coordinates.}
	\label{tab:temp-stability-rand}
\end{table}

\begin{result}
	The clusterings computed with K-Means and Spectral Clustering are not temporally stable. A significant number of nodes would be assigned to different clusters in different years. 
\end{result}

We notice that Rand indices fluctuate considerably when different pairs of years are considered and lie in the intervals [0.57, 0.95] for $k=2$, [0.48, 0.95] for $k=3$ and [0.44, 0.86] for $k=4$. This suggests that clusterings are not temporally stable. 
The price zone configuration identified for a specific year might be far from the optimal configuration corresponding to a different year. This challenges the goal of finding price zones that reflect stable long-term patterns in nodal prices.

\begin{figure}[!htp]
    \begin{subfigure}{0.5\textwidth}
        \centering
        \includegraphics[width=\linewidth]{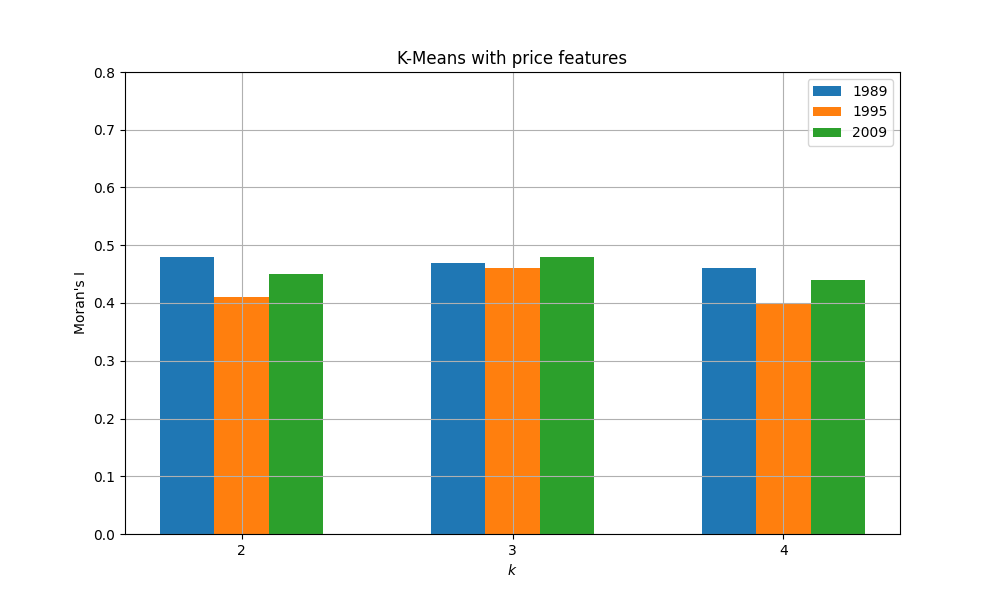}
    \end{subfigure}
    \begin{subfigure}{0.5\textwidth}
        \centering
        \includegraphics[width=\linewidth]{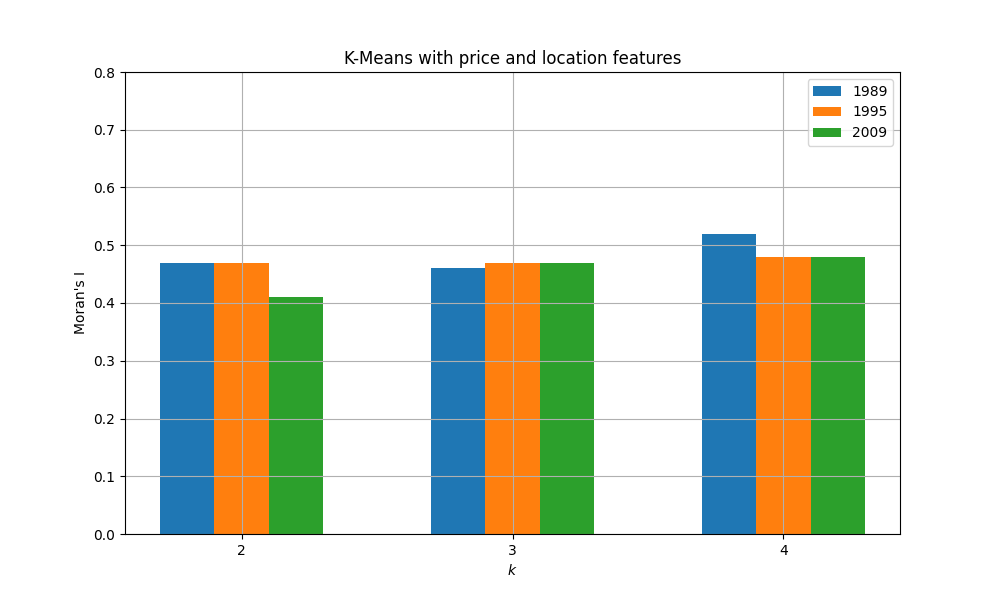}
    \end{subfigure}\\[1ex]
    \begin{subfigure}{0.5\textwidth}
        \centering
        \includegraphics[width=\linewidth]{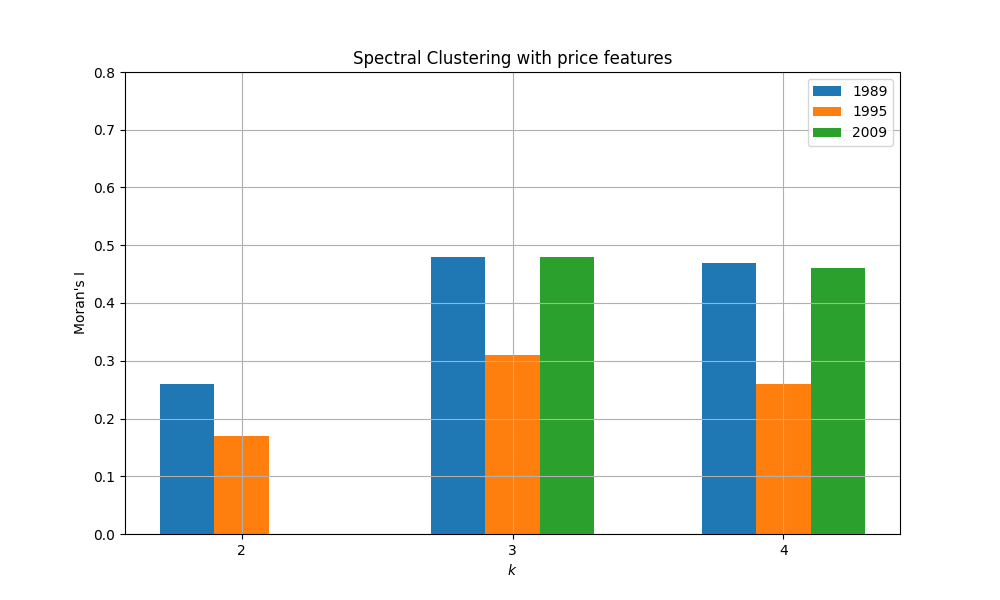}
    \end{subfigure}
    \begin{subfigure}{0.5\textwidth}
        \centering
        \includegraphics[width=\linewidth]{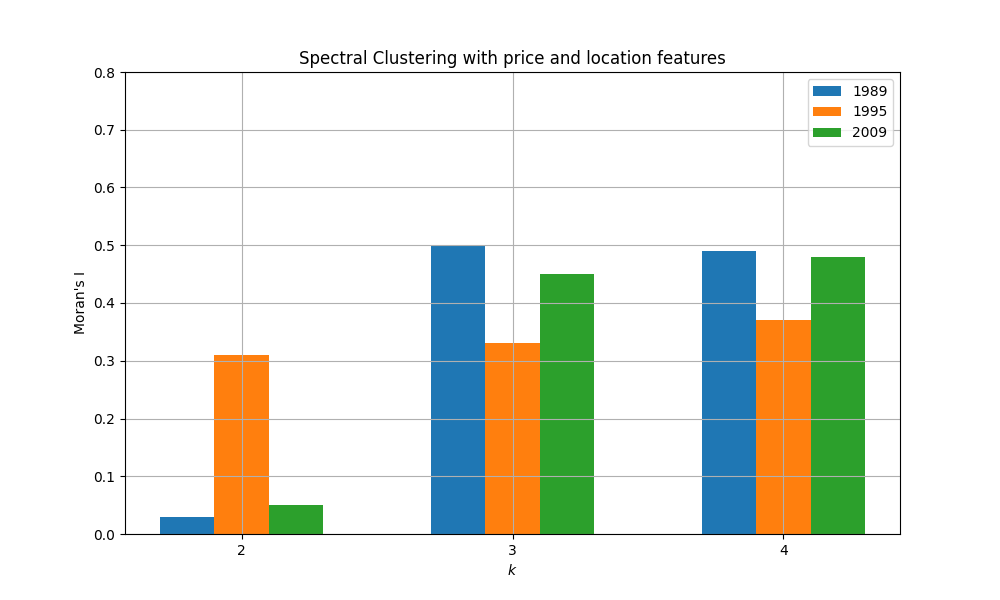}
    \end{subfigure}
    \caption{Global Moran's I values of the clusterings computed with K-Means and Spectral Clustering}
    \label{fig:moran}
\end{figure}

Last but not least, we analyze how spatially coherent the clusterings computed with K-Means and Spectral Clustering for \textcolor{black}{the scenario years} 1989, 1995 and 2009 are.
Figure \ref{fig:moran} illustrates the Global Moran's I values of the clusterings computed using (1) nodal weekly prices and (2) nodal weekly prices and the location coordinates of nodes as features.

For all years we consider, the Global Moran's I values are between 0.01 and 0.5, indicating that clusterings exhibit moderate positive spatial autocorrelation.
Neighboring nodes are more likely to have similar prices and to belong to the same cluster.
Also, considering the location coordinates in the K-Means computation leads to better Global Moran's I scores which are quite stable across time.
However, the plots from Figure \ref{fig:moran} show that the clusterings obtained with Spectral Clustering exhibit scores that vary considerably across years. 

Analyzing the clusters on the map of Germany, we notice that isolated points assigned to a specific cluster are sometimes located in the geographic area predominantly occupied by points belonging to a different cluster.
This behavior can also be observed when latitude and longitude are included as features and is independent of the number of clusters or clustering algorithm. 
Therefore, a post-processing step is required to obtain spatially coherent price zones. This is because prices alone do not define geographically well-separated or spatially coherent clusters.
This phenomenon can be observed in Figures \ref{fig:rand-vis-4-clusters-sc} and \ref{fig:rand-vis-4-clusters-kmeans} from Appendix \ref{app:rand-visualization}.

\subsection*{Sensitivity Analysis regarding Outliers}
In what follows, we briefly discuss the influence of outliers on the resulting price zones. 
We classify a node as an outlier if it has a price lower than 0 EUR/MWh or larger than 100 EUR/MWh in any of the periods considered in the clustering computation.
Out of 1,478,400 nodal prices reported for a specific year, 38,378 ($\approx 2.59 \%$) lie outside the interval $[0, 100]$.
Interestingly, this number is identical in all three years we consider. 
We want to understand the influence of outliers on the clustering computation. To do this, we clip the prices to the interval $[0, 100]$ and apply the clustering algorithms.
Table \ref{tab:mean-prices-without-outliers} contains the average yearly prices of the clusters computed with K-Means and Spectral Clustering after clipping the prices.
\begin{table}[!htp]
	\centering
	\begin{tabular}{c|c|c|c|c|c|c|c||c|c|c|c|c}
		& & & \multicolumn{5}{c||}{K-Means} & \multicolumn{5}{c}{Spectral} \\
		Year & $k$ & Loc. & $\Delta_\mu$ & $\mu_1$ & $\mu_2$ & $\mu_3$ & $\mu_4$ & $\Delta_\mu$ & $\mu_1$ & $\mu_2$ & $\mu_3$ & $\mu_4$ \\
		\hline 
		1989 & 2 & & 5.7 & 46.59 & 40.89 & - & - & 3.2 & 43.66 & 46.86 & - & - \\
		1995 & 2 & & 5.89 & 41.54 & 47.43 & - & - & 3.2 & 43.66 & 46.86 & - & - \\
		2009 & 2 & & 5.9 & 47.44 & 41.54 & - & - & 3.2 & 43.66 & 46.86 & - & -\\
		\hline 
		1989 & 2 & x & 5.79 & 41.25 & 47.04 & - & - & 3.13 & 43.63 & 46.76 & - & - \\
		1995 & 2 & x & 5.68 & 46.55 & 40.87 & - & - & 3.13 & 43.63 & 46.76 & - & - \\
		2009 & 2 & x & 5.79 & 41.32 & 47.11 & 3.13 & 43.63 & 46.76 & - & - \\
		\hline
		1989 & 3 &  & 3.95 & 41.52 & 41.47 & 47.39 & - & 2.21 & 43.39 & 46.64 & 46.7 & -  \\
		1995 & 3 &  & 4.27 & 41.2 & 47.6 & 46.04 & - & 3.67 & 43.94 & 46.9 & 41.4 & -  \\
		2009 & 3 &  & 4.44 & 40.91 & 47.57 & 44.16 & - & 3.76 & 43.96 & 41.26 & 46.9 & -  \\
		\hline 
		1989 & 3 & x & 4.75 & 42.43 & 47.38 & 40.26 & - & 3.71 & 43.91 & 46.85 & 41.29 & - \\
		1995 & 3 & x & 5.06 & 41.36 & 46.72 & 48.95 & - & 3.71 & 43.91 & 46.85 & 41.29 & - \\
		2009 & 3 & x & 5.02 & 48.61 & 41.08 & 46.43 & - & 3.71 & 43.91 & 46.85 & 41.29 & - \\
		\hline 
		1989 & 4 & & 3.96 & 47.1 & 41.02 & 41.76 & 39.43 & 3.04 & 43.74 & 46.28 & 46.57 & 41.33 \\
		1995 & 4 & & 4.7 & 40.72 & 43.74 & 48.92 & 47.32 & 2.86 & 40.94 & 46.27 & 46.58 & 46.51 \\
		2009 & 4 & & 3.94 & 41.34 & 47.48 & 40.24 & 43.25 & 3.07 & 43.72 & 46.53 & 41.33 & 46.54 \\
		\hline 
		1989 & 4 & x & 4.2 & 41.21 & 47.48 & 43.73 & 39.92 & 1.85 & 43.16 & 46.64 & 46.61 & 45.94 \\
		1995 & 4 & x & 3.82 & 40.62 & 43.89 & 41.62 & 47.5 & 1.79 & 42.99 & 46.21 & 46.25 & 46.56 \\
		2009 & 4 & x & 4.5 & 41.51 & 47.69 & 45.65 & 40.06 & 1.85 & 43.16 & 46.64 & 46.61 & 45.94 \\
	\end{tabular}
	\caption{Cluster average prices after clipping nodal prices to $[0, 100]$ (EUR/MWh)}
	\label{tab:mean-prices-without-outliers}
\end{table}
We observe that these prices are similar to the average yearly prices shown in Table \ref{tab:1}, in which clusterings were computed without clipping prices.
The global average distances between cluster mean prices (i.e., $\Delta_\mu$) illustrated in Table \ref{tab:mean-prices-without-outliers} are also close to the average distances included in Table \ref{tab:1}. In particular, the differences between the values of $\Delta_\mu$ shown in these tables are smaller than 6 EUR/MWh.
Hence, the impact of outliers on the resulting clusters or price zones is not considerable. 

\color{black}
\subsection{Effects of unstable zonal configurations} 
Our experiments show that the alternative bidding zone configurations for Germany exhibit low price similarity within zones and high price similarity between zones, and are not stable over time. 
Unstable configurations with respect to these criteria have negative implications related to equity and market efficiency.

\begin{figure}[!htp]
	\centering
    \begin{subfigure}{0.45\textwidth}
        \centering
        \includegraphics[width=\linewidth]{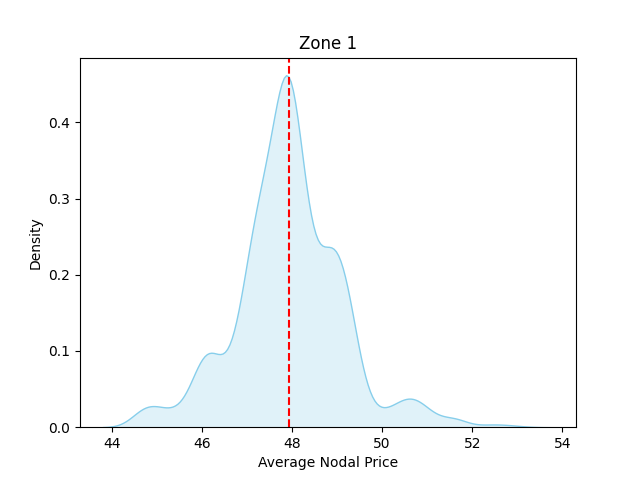}
    \end{subfigure}
    \begin{subfigure}{0.45\textwidth}
        \centering
        \includegraphics[width=\linewidth]{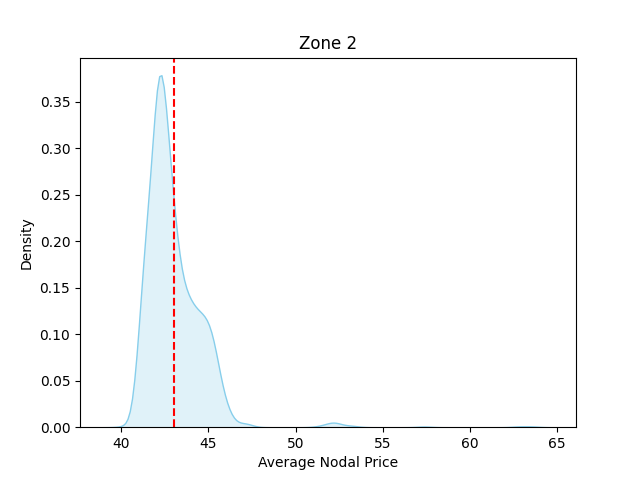}
    \end{subfigure}\\[1ex]
    \caption{Density plots showing the distribution of average nodal prices (EUR/MWh) in the two zones of the DE2 (k-means) configuration, with the average zonal prices highlighted for comparison}
    \label{fig:effects}
\end{figure}

\textcolor{black}{The density plots from Figure \ref{fig:effects} illustrate the distribution of average nodal prices in the two zones of the DE2 (k-means) configuration, with the average zonal prices highlighted for comparison. The average nodal prices were computed considering all prices from the LMP Study (see Section \ref{sec:data}). We observe that the distribution corresponding to zone 1 has long tails, which indicates large variations in nodal prices within zones. These variations}
lead to regional inequities and distort the zonal price signals, meaning the prices do not accurately reflect the supply and demand conditions in specific subregions. This results in poor local incentives to adapt to energy prices, poor investment incentives, and an inefficient implementation of scarcity pricing \citep{papavasiliou2020scarcity}.
\textcolor{black}{Furthermore, the distribution for zone 2 reveals that over 50\% of the nodes in this zone have average prices below the average zonal price. If the uniform price for this zone was set at the average zonal price, it would create equity issues among the market agents located at these nodes.}
\color{black}

\color{black}
Our analysis suggests that nodes with similar prices are assigned to different zones under the alternative configurations being discussed \textcolor{black}{(see Appendix \ref{app:cluster-price-analysis})}. This discrepancy can create economic inefficiencies and fairness issues if consumers and producers in these zones experience different electricity prices \citep{morales2018efficiency}.

Morevoer, temporally unstable zonal configurations \textcolor{black}{(Table \ref{tab:compare-official-bz-rand})} can lead to long-term market inefficiencies by distorting price signals, reducing market liquidity, and creating investment uncertainty \citep{gugler2020investment}. Inconsistent price patterns over time can diminish market liquidity by introducing uncertainty and disrupting trading activities. Moreover, this uncertainty about future market conditions can discourage investment in new generation capacity and infrastructure upgrades.
\color{black}

\section{Conclusions} \label{sec:conclusion}
In this paper, we analyzed the alternative price zone configurations proposed by ACER for Germany, namely DE2 (k-means), DE2 (spectral), DE3 (spectral) and DE4 (spectral). 
Similar to ACER's approach (the algorithms used are publicly known, but the implementation details are not), we computed alternative price zone configurations by solving a clustering problem using K-Means and Spectral Clustering. 
Clusterings that we obtained with K-Means using only latitude and longitude as features are similar to the configurations proposed by ACER. 
Additionally, we computed alternative configurations by taking nodal prices or nodal prices and geographic coordinates into account and compared them with the configurations proposed by ACER. 
The resulting price zones were different from the proposed ACER configurations. 
Moreover, our results show that clusters with ``minimal price dispersion within each bidding zone'' \citep{acer-list-all-proposed-bzs} were not geographically coherent on the map, and they do not have similar numbers of nodes. 
Note that ACER might have had access to a more complete dataset such that a one-to-one comparison to their results is not possible. We can only leverage the data that was made available publicly. 

We evaluated the stability of the alternative configurations considering three criteria, namely intra- and inter-cluster similarity, temporal stability and spatial coherence. Our findings have important implications concerning the ongoing discussion of whether a split of the German bidding zone should be performed. 

Firstly, splitting the German bidding zone into 2, 3 or 4 zones does not reduce the price dispersion within zones by more than 3-9\%. 
Moreover, the decrease in average price standard deviation is not monotonic with respect to the number of zones. Configurations with 3 or 4 zones do not necessarily reduce the average price standard deviation compared to configurations with fewer zones.
Furthermore, for all proposed configurations, the average prices across zones do not differ considerably. The annual differences between the average cluster prices are less than 6 EUR/MWh for all three years we analyzed. 
As discussed in \citep{acer-methodology}, two main goals of a split are to reduce price dispersion within individual bidding zones and to reflect long-term differences in average prices across the resulting zones. Our analysis shows that the alternative configurations proposed by ACER can be challenged based on these objectives.

Secondly, independent of how the alternative price zone configurations were determined and whether we look at those suggested by ACER or our own, none of them is temporally stable with respect to prices. Recall that prices were the only clustering feature in ACER's methodology for obtaining alternative configurations \citep{acer-clustering-algo} and ``\textit{can be used as proxies for economic efficiency}'' \citep{acer-list-all-proposed-bzs}. Our study shows that a configuration that is optimal in one time period might turn out to be suboptimal in another. 
There can be many arguments for or against a bidding zone split. However, when a decision should be made regarding the configuration of Germany into bidding zones, the data revealed and the method defined during the BZR do not allow for a clear recommendation.

\section*{Data Availability}
The data used for this publication are available on the Bidding Zone Review website \citep{bz-website}.

\section*{Acknowledgements}
We gratefully acknowledge the financial support of the Kopernikus-Project ``SynErgie" by the Federal Ministry of Education and Research of Germany (BMBF) and the project supervision by the project management organization Projekttr\"ager J\"ulich (PtJ).

\bibliographystyle{apalike-ejor}
\bibliography{bibliography} 

\begin{appendices}
	\section{Approximations of the alternative configurations proposed by ACER} \label{app:approx-bz-acer}

\begin{figure}[h]
    \centering
    \includegraphics[scale=0.72]{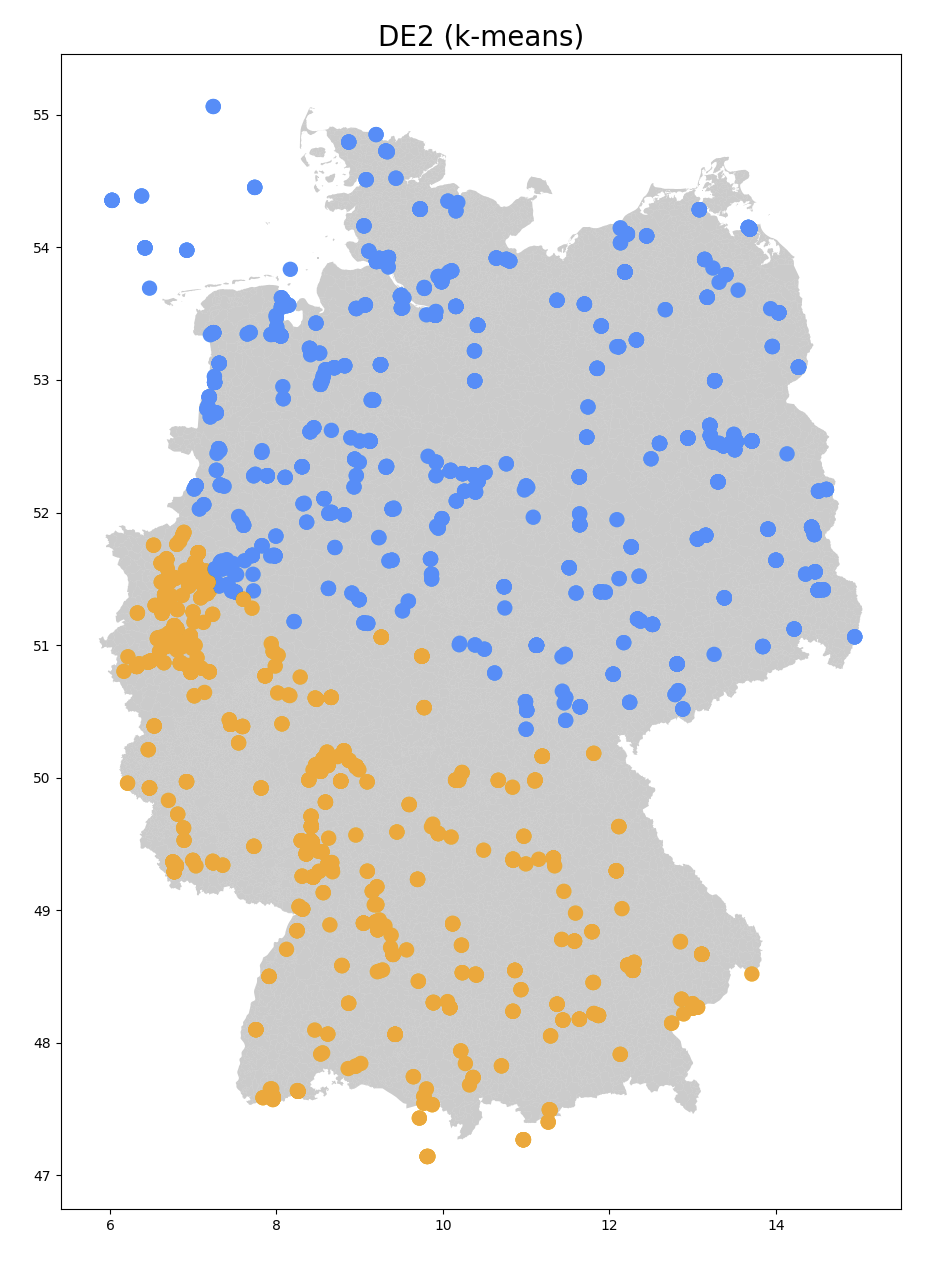}
    \includegraphics[scale=0.72]{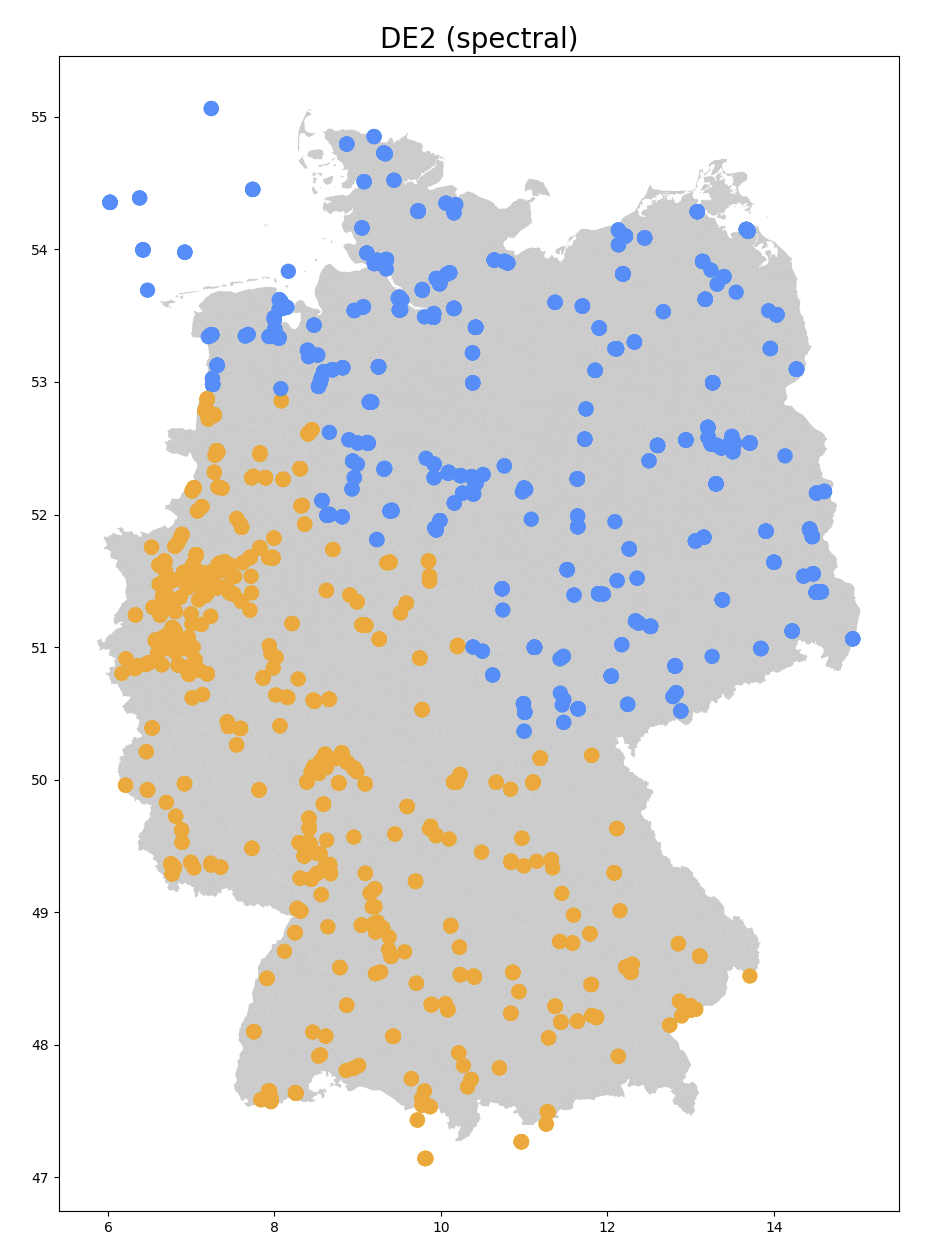} \\
    
    \includegraphics[scale=0.72]{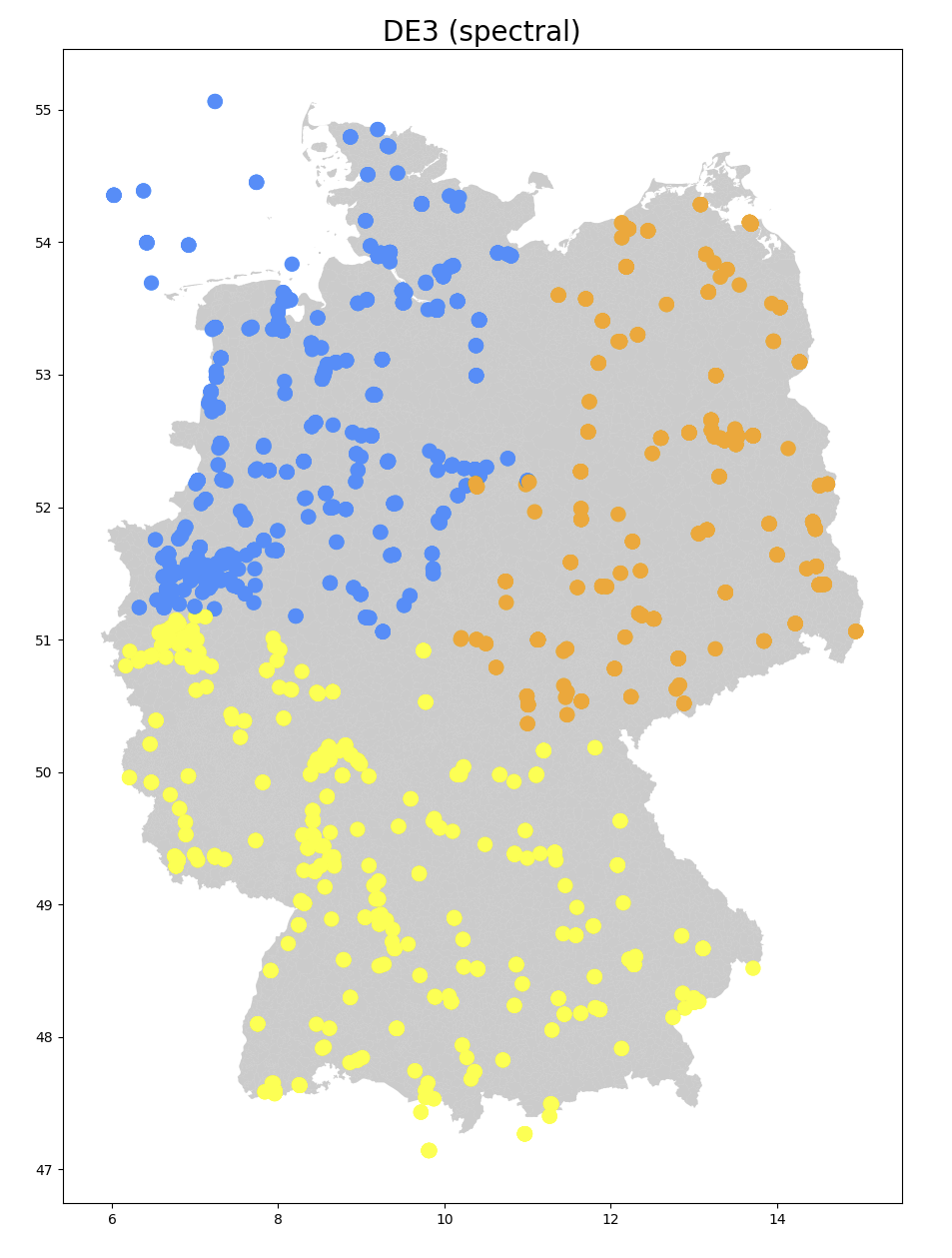}
    \includegraphics[scale=0.72]{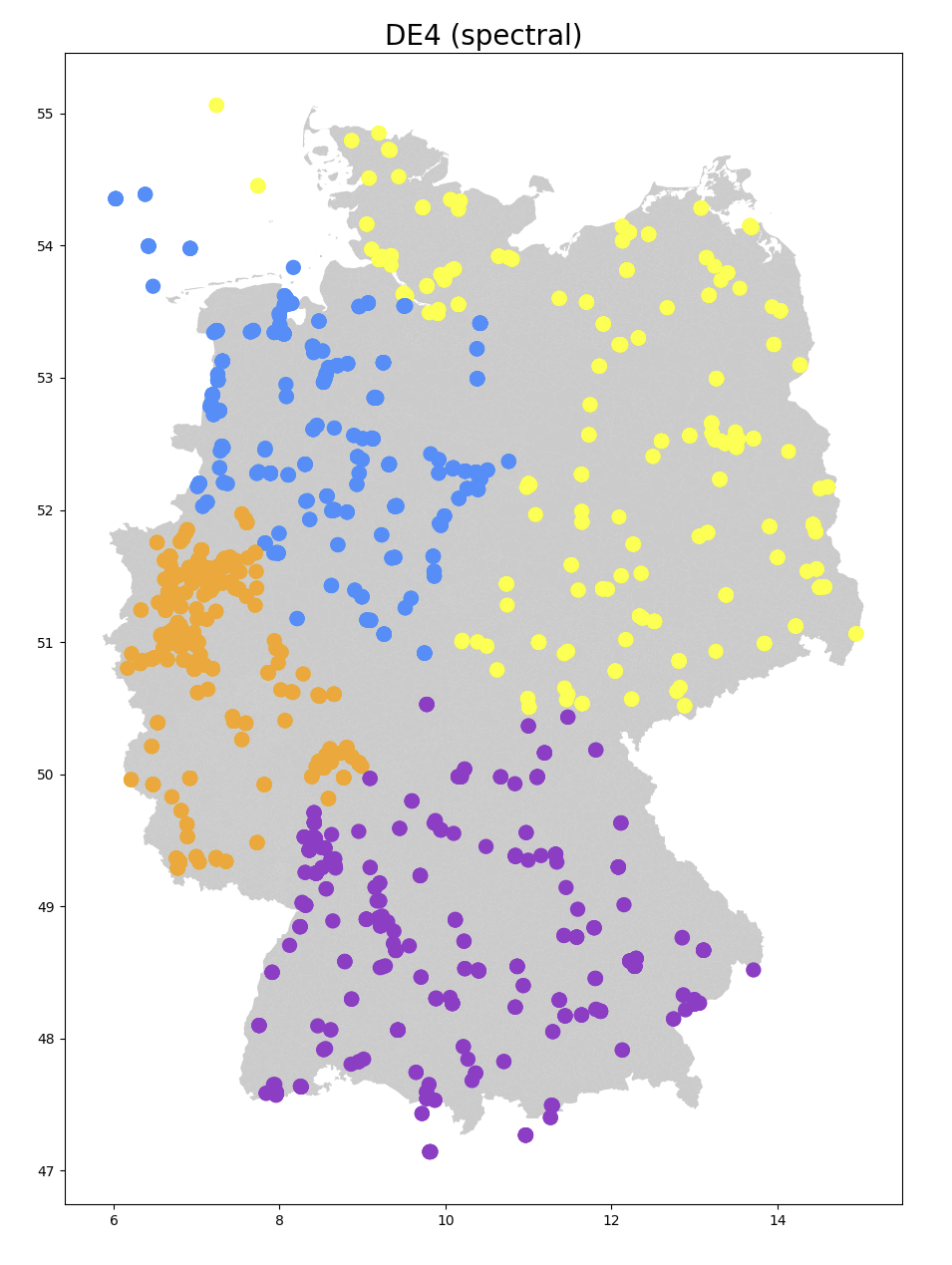}
    
    \caption{Approximations of the alternative configurations proposed by ACER}
    \label{fig:proposed-zones}
\end{figure}

\newpage

	\section{Cluster price analysis for the alternative configurations proposed by ACER} \label{app:cluster-price-analysis}

\subsection{Cluster price standard deviations} \label{app:cluster-variances}

Table \ref{tab:price-std-summary} shows the average price standard deviations of a single price zone configuration as well as of the configurations proposed by ACER.

\begin{table}[!htp]
    \centering
    \begin{tabular}{c|c|c|c}
        & 1989 & 1995 & 2009 \\
        \hline
        Single price zone & 14.54 & 14.32 & 11.48 \\
        DE2 (k-means) & 13.09 & 12.93 & 10.85 \\ 
        DE2 (spectral) & 13.98 & 13.80 & 11.22 \\
        DE3 (spectral) & 13.86 & 13.65 & 11.26 \\
        DE4 (spectral) & 13.14 & 13.02 & 10.87 \\
    \end{tabular}
    \caption{Average price standard deviations of configurations with 1, 2, 3 or 4 price zones}
    \label{tab:price-std-summary}
\end{table}

\begin{figure}[!htp]
	\centering
	\includegraphics[width=.32\textwidth]{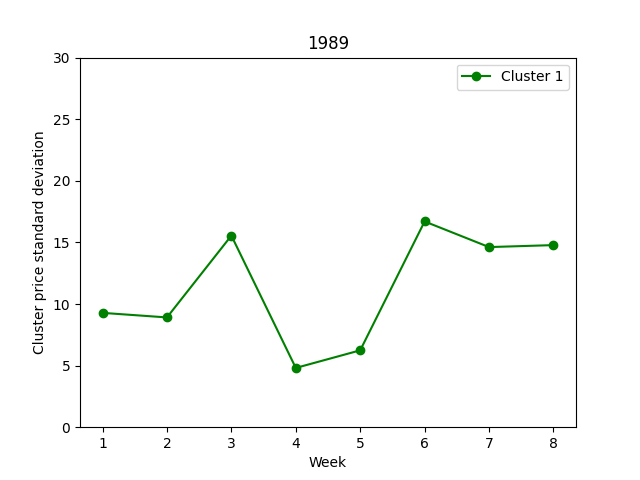}
	\includegraphics[width=.32\textwidth]{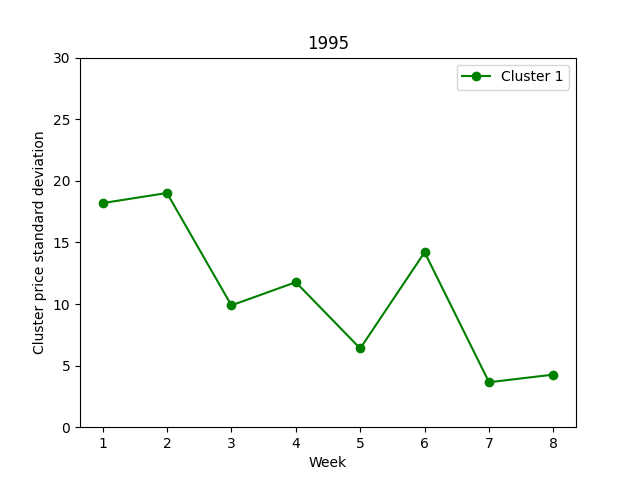}
	\includegraphics[width=.32\textwidth]{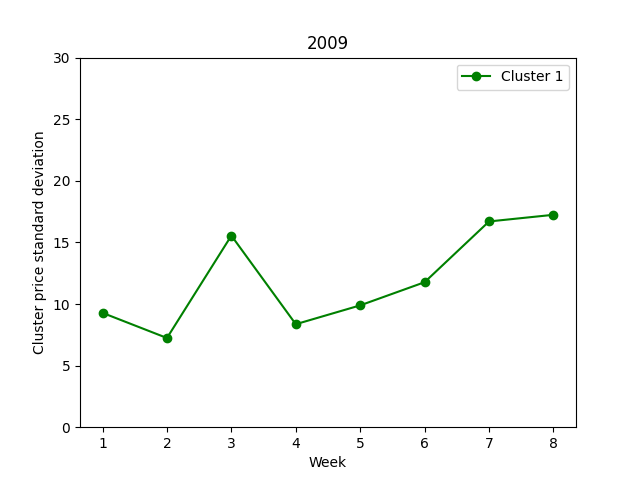}
	\caption{Single price zone (cluster)}
	\label{fig:1-zone-price-var}

	\centering
	\includegraphics[width=.32\textwidth]{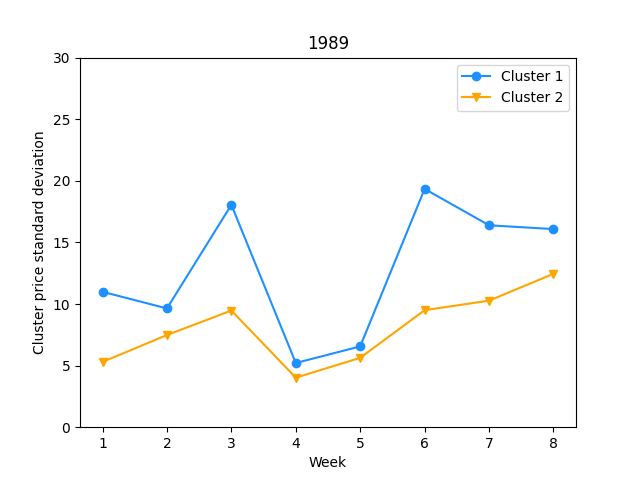}
	\includegraphics[width=.32\textwidth]{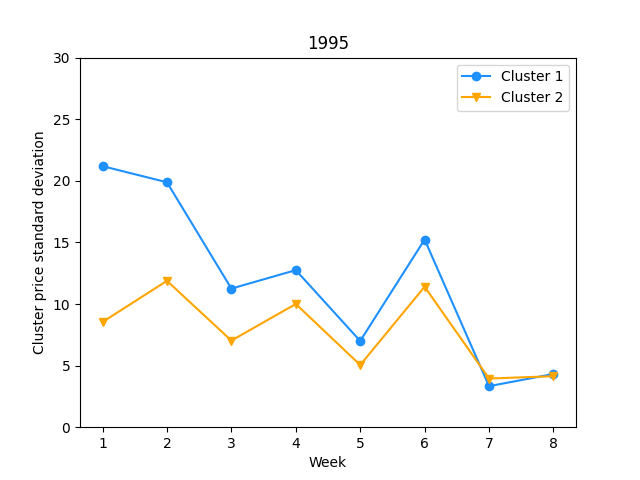}
	\includegraphics[width=.32\textwidth]{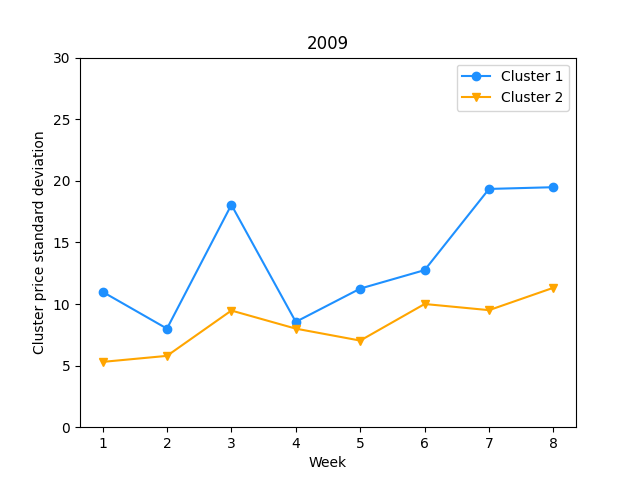}
	\caption{DE2 (k-means)}
	\label{fig:de2-kmeans-price-var}
\end{figure}

The following plots illustrate, for every configuration, the cluster price standard deviation corresponding to each week in \textcolor{black}{the scenario years} 1989, 1995, and 2009. 
A point represents the price standard deviation of a specific cluster in a specific week. Depending on the configuration (i.e., number of clusters), multiple points are shown for the same week.
Figure \ref{fig:1-zone-price-var} highlights the weekly price standard deviation of a single price zone for each year we consider.

\begin{figure}[!htp]
	\centering
	\includegraphics[width=.32\textwidth]{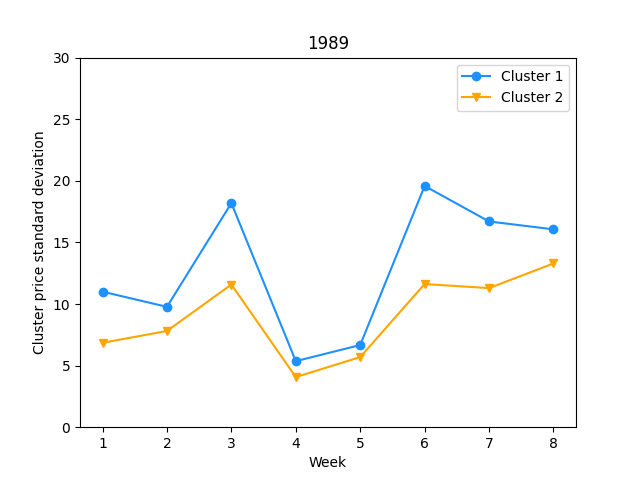}
	\includegraphics[width=.32\textwidth]{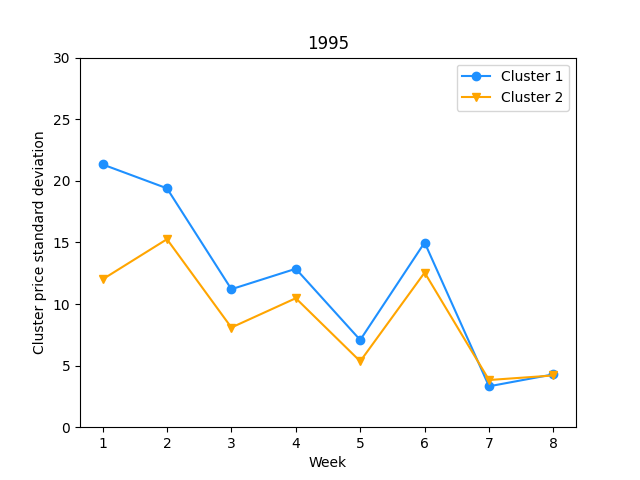}
	\includegraphics[width=.32\textwidth]{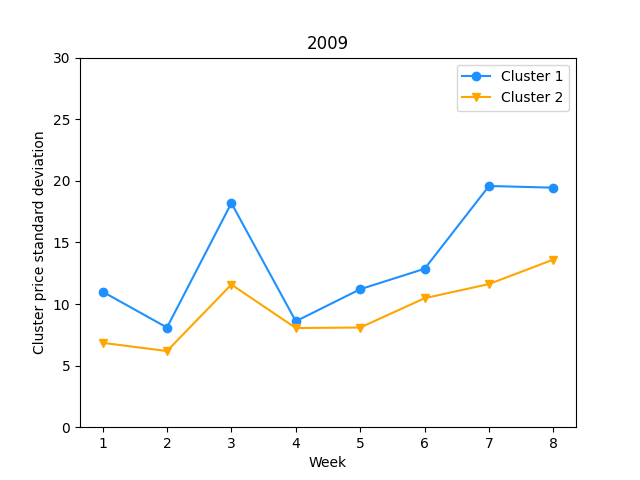}
	\caption{DE2 (spectral)}
	\label{fig:de2-spectral-price-var}

	\centering
	\includegraphics[width=.32\textwidth]{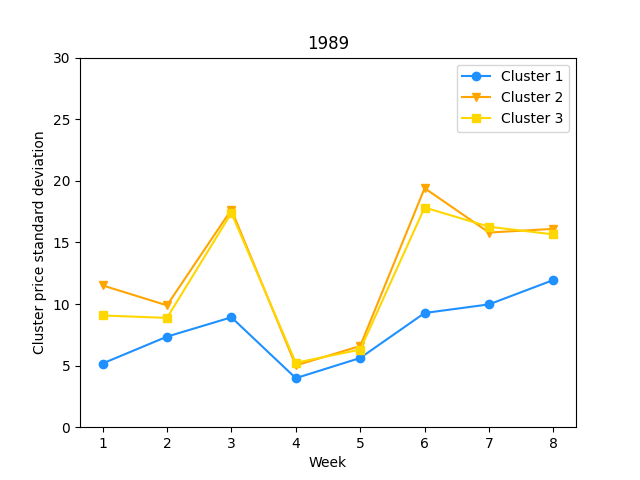}
	\includegraphics[width=.32\textwidth]{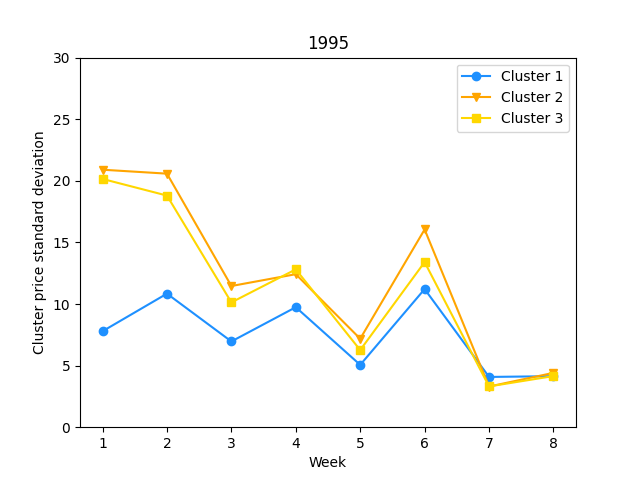}
	\includegraphics[width=.32\textwidth]{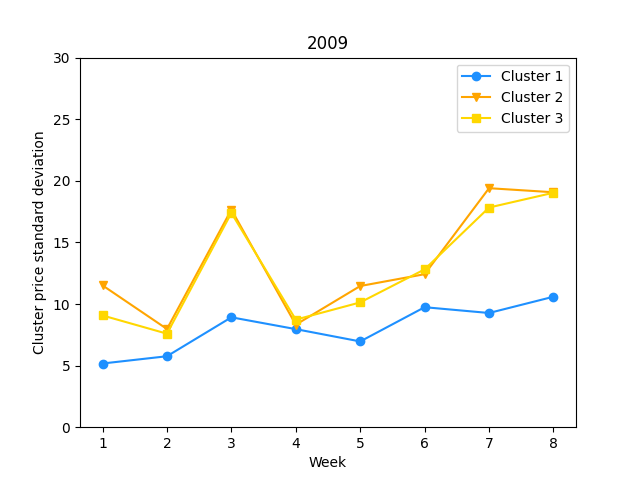}
	\caption{DE3 (spectral)}
	\label{fig:de3-spectral-price-var}

	
    \centering
	\includegraphics[width=.32\textwidth]{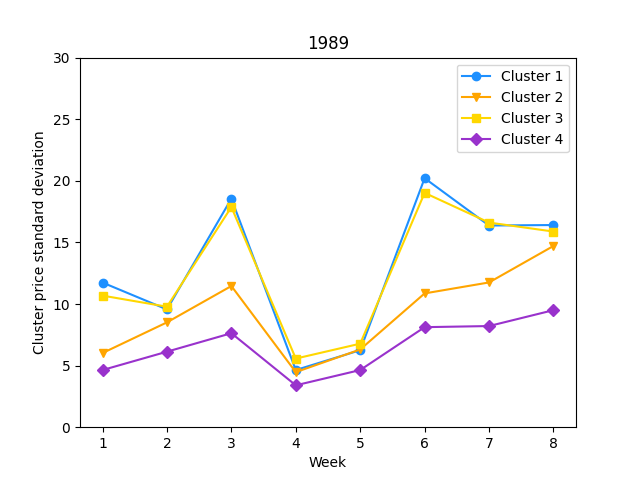}
	\includegraphics[width=.32\textwidth]{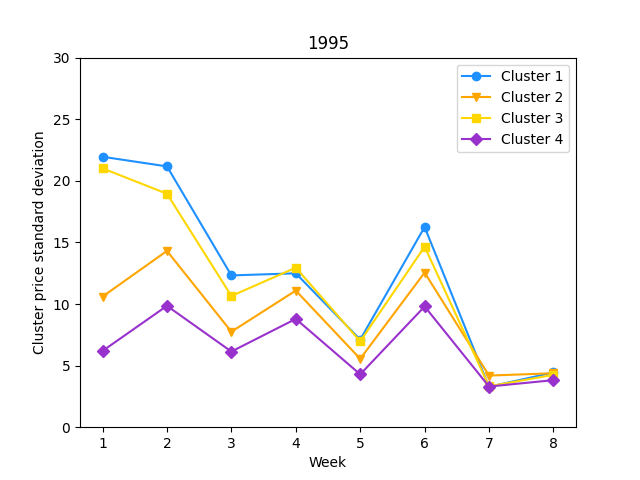}
	\includegraphics[width=.32\textwidth]{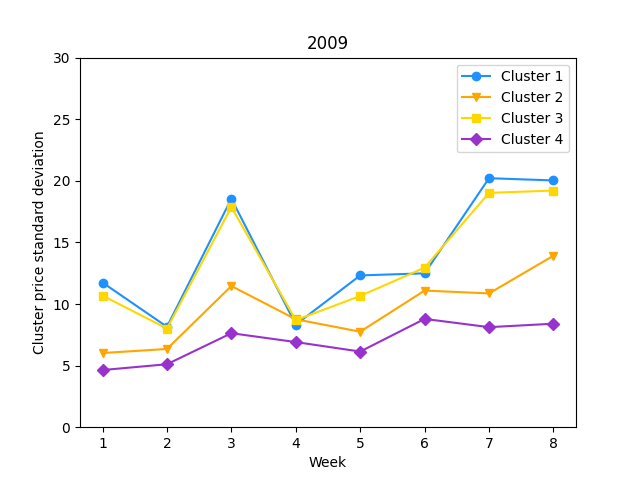}
	\caption{DE4 (spectral)}
	\label{fig:de4-spectral-price-var}
\end{figure}

\newpage 

\subsection{Average cluster prices} \label{app:cluster-average-prices}

In the following tables, $\Delta^{\text{year}}_\mu$ represents the average distance between cluster mean prices for a specific year, i.e., $\Delta^{\text{year}}_\mu = \frac{1}{8} \displaystyle\sum_{i \in \{1, \dots, 8\}} \Delta^{\text{W}_i}_\mu$, where $\Delta^{\text{W}_i}_\mu$ is the average distance between cluster mean prices for week $\text{W}_i$ (see Section \ref{sec:intra-inter-similarity}).
\begin{table}[!htp]
    \centering
    \begin{tabular}{c|c|c|c|c|c|c|c|c|c||c}
        & & W1 & W2 & W3 & W4 & W5 & W6 & W7 & W8 & $\Delta^{\text{year}}$ \\
        \hline
        1989 & Cluster 1 & 50.49 & 46.72 & 41.99 & 49.88 & 48.52 & 39.52 & 40.22 & 45.50 & - \\
        1995 & Cluster 1 & 39.87 & 28.09 & 45.89 & 46.11 & 48.61 & 38.35 & 51.76 & 54.50 & - \\
        2009 & Cluster 1 & 50.49 & 51.94 & 41.99 & 47.66 & 45.89 & 46.11 & 39.52 & 40.91 & - \\
    \end{tabular}
    \caption{Cluster average prices (EUR/MWh) - Single price zone (cluster)}
    \label{tab:average-prices-1-cluster}
 \end{table}

 \begin{table}[!htp]
    \centering
    \begin{tabular}{c|c|c|c|c|c|c|c|c|c||c}
        & & W1 & W2 & W3 & W4 & W5 & W6 & W7 & W8 & $\Delta^{\text{year}}_\mu$ \\
        \hline
        \multirow{2}{*}{1989} & Cluster 1 & 49.09 & 45.71 & 39.18 & 49.32 & 47.94 & 35.84 & 37.44 & 44.38 & \multirow{2}{*}{4.3}  \\
        & Cluster 2 & 52.55 & 48.21 & 46.10 & 50.69 & 49.39 & 44.91 & 44.28 & 47.15 \\
        \hline

        \multirow{2}{*}{1995} & Cluster 1 & 35.02 & 21.21 & 44.56 & 45.13 & 47.65 & 35.63 & 51.22 & 54.14 & \multirow{2}{*}{5.75} \\
        & Cluster 2 & 46.98 & 38.19 & 47.84 & 47.55 & 50.03 & 42.35 & 52.56 & 55.04 \\
        \hline

        \multirow{2}{*}{2009} & Cluster 1 & 49.09 & 51.21 & 39.18 & 47.11 & 44.56 & 45.13 & 35.84 & 37.25 & \multirow{2}{*}{4.67} \\
        & Cluster 2 & 52.55 & 53.01 & 46.10 & 48.46 & 47.84 & 47.55 & 44.91 & 46.28 \\
    
    \end{tabular}
    \caption{Cluster average prices (EUR/MWh) - DE2 (k-means)}
    \label{tab:average-prices-DE2-kmeans}
 \end{table}

\begin{table}[!htp]
    \centering
    \begin{tabular}{c|c|c|c|c|c|c|c|c|c||c}
        & & W1 & W2 & W3 & W4 & W5 & W6 & W7 & W8 & $\Delta^{\text{year}}_\mu$ \\
        \hline
        \multirow{2}{*}{1989} & Cluster 1 & 48.87 & 45.52 & 38.77 & 49.13 & 47.73 & 34.89 & 36.80 & 44.29 &  \multirow{2}{*}{4.17} \\
        & Cluster 2 & 52.07 & 47.90 & 45.14 & 50.61 & 49.31 & 44.05 & 43.56 & 46.70 \\
        \hline

        \multirow{2}{*}{1995} & Cluster 1 & 34.06 & 20.40 & 44.22 & 44.91 & 47.36 & 35.03 & 51.08 & 54.08 &  \multirow{2}{*}{5.46} \\
        & Cluster 2 & 45.56 & 35.63 & 47.53 & 47.29 & 49.84 & 41.61 & 52.44 & 54.91 \\
        \hline

        \multirow{2}{*}{2009} & Cluster 1 & 48.87 & 51.06 & 38.77 & 46.91 & 44.22 & 44.91 & 34.89 & 36.81 &  \multirow{2}{*}{4.47} \\
        & Cluster 2 & 52.07 & 52.80 & 45.14 & 48.39 & 47.53 & 47.29 & 44.05 & 44.93 \\
    \end{tabular}
    \caption{Cluster average prices (EUR/MWh) - DE2 (spectral)}
    \label{tab:average-prices-DE2-spectral}
 \end{table}

 \begin{table}[!htp]
    \centering
    \begin{tabular}{c|c|c|c|c|c|c|c|c|c||c}
        & & W1 & W2 & W3 & W4 & W5 & W6 & W7 & W8 & $\Delta^{\text{year}}_\mu$  \\
        \hline
        \multirow{2}{*}{1989} & Cluster 1 & 52.68 & 48.38 & 46.38 & 50.72 & 49.36 & 44.96 & 44.58 & 47.43 &  \multirow{3}{*}{3.09} \\
        & Cluster 2 & 49.17 & 45.94 & 40.29 & 49.85 & 48.51 & 37.41 & 38.86 & 44.68 \\
        & Cluster 3 & 49.72 & 45.81 & 39.04 & 48.88 & 47.51 & 35.88 & 36.81 & 44.32 \\
        \hline

        \multirow{2}{*}{1995} & Cluster 1 & 47.30 & 38.88 & 47.80 & 47.66 & 50.05 & 42.57 & 52.58 & 55.08 & \multirow{3}{*}{3.99} \\
        & Cluster 2 & 36.69 & 23.49 & 45.31 & 45.88 & 48.37 & 36.26 & 51.77 & 54.24 \\
        & Cluster 3 & 35.34 & 21.51 & 44.37 & 44.53 & 47.18 & 36.21 & 50.73 & 54.18 \\
        \hline
        
        \multirow{2}{*}{2009} & Cluster 1 & 52.68 & 53.05 & 46.38 & 48.42 & 47.8 & 47.66 & 44.96 & 46.74 & \multirow{3}{*}{3.37} \\
        & Cluster 2 & 49.17 & 51.49 & 40.29 & 47.89 & 45.31 & 45.88 & 37.41 & 38.87 \\
        & Cluster 3 & 49.72 & 51.21 & 39.04 & 46.38 & 44.37 & 44.53 & 35.88 & 36.69 \\
    \end{tabular}
    \caption{Cluster average prices (EUR/MWh) - DE3 (spectral)}
    \label{tab:average-prices-DE3-spectral}

\vspace{1.5cm}

    \centering
    \begin{tabular}{c|c|c|c|c|c|c|c|c|c||c}
        & & W1 & W2 & W3 & W4 & W5 & W6 & W7 & W8 & $\Delta^{\text{year}}_\mu$  \\
        \hline
        \multirow{4}{*}{1989} & Cluster 1 & 49.03 & 46.0 & 39.51 & 49.91 & 48.52 & 36.68 & 38.28 & 44.48 & \multirow{4}{*}{3.14} \\
        & Cluster 2 & 52.30 & 47.85 & 45.41 & 50.94 & 49.77 & 44.74 & 43.06 & 46.37 \\
        & Cluster 3 & 49.00 & 45.41 & 38.73 & 48.84 & 47.44 & 34.78 & 36.55 & 44.25 \\
        & Cluster 4 & 52.62 & 48.50 & 46.44 & 50.37 & 48.91 & 44.81 & 45.46 & 47.86 \\
        \hline
        
        \multirow{4}{*}{1995} & Cluster 1 & 35.24 & 21.24 & 44.78 & 45.75 & 48.32 & 36.11 & 51.63 & 54.13 & \multirow{4}{*}{4.19} \\
        & Cluster 2 & 45.35 & 35.98 & 47.97 & 47.80 & 50.53 & 41.13 & 53.33 & 55.07 \\
        & Cluster 3 & 34.37 & 20.74 & 44.23 & 44.60 & 47.05 & 34.99 & 50.87 & 54.11 \\
        & Cluster 4 & 48.32 & 39.51 & 47.58 & 47.13 & 49.35 & 43.53 & 51.58 & 54.94 \\
        \hline
        
        \multirow{4}{*}{2009} & Cluster 1 & 49.03 & 51.29 & 39.51 & 47.72 & 44.78 & 45.75 & 36.68 & 37.71 & \multirow{4}{*}{3.34} \\
         & Cluster 2 & 52.30 & 53.05 & 45.41 & 49.11 & 47.97 & 47.80 & 44.74 & 44.95 \\
         & Cluster 3 & 49.00 & 51.07 & 38.73 & 46.6 & 44.23 & 44.60 & 34.78 & 36.69 \\
         & Cluster 4 & 52.62 & 52.87 & 46.44 & 47.67 & 47.58 & 47.13 & 44.81 & 47.18 \\
    \end{tabular}
    \caption{Cluster average prices - DE4 (spectral)}
    \label{tab:average-prices-DE4-spectral}
 \end{table}

	\section{Intra- and inter-cluster similarity evaluation} \label{app:intra-inter}

\begin{table}[!htp]
	\centering
	\begin{tabular}{c|c|c|c|c|c|c|c|c|c|c|c|c}
		Year & $k$ & Alg. & Loc. & $\sigma$ & $\Delta_\mu$ & $\Delta_\text{min}$ & $\Delta_\text{avg}$ & $DB$ & $\mu_1$ & $\mu_2$ & $\mu_3$ & $\mu_4$ \\
		\hline 
		1989 & 2 & K-Means &  & 13.4 & 5.91 & 0.0 & 5.91 & 0.63 & 47.28 & 41.37 & - & - \\
		1995 & 2 & K-Means &  & 11.16 & 4.66 & 0.0 & 4.97 & 0.76 & 48.47 & 43.81 & - & - \\
		2009 & 2 & K-Means &  & 10.8 & 3.68 & 0.0 & 3.68 & 0.71 & 45.29 & 48.97 & - & - \\
		\hline 
		1989 & 2 & Spectral &  & 13.39 & 3.57 & 0.0 & 4.3 & 1.44 & 43.61 & 47.18 & - & - \\
		1995 & 2 & Spectral &  & 15.16 & 2.87 & 0.0 & 3.57 & 1.89 & 44.74 & 41.87 & - & - \\
		2009 & 2 & Spectral &  & 11.61 & 0.31 & 0.0 & 2.19 & 13.76 & 46.68 & 46.37 & - & - \\
		\hline 
		1989 & 2 & K-Means & x & 13.0 & 5.91 & 0.0 & 5.93 & 0.62 & 41.64 & 47.55 & - & - \\
		1995 & 2 & K-Means & x & 13.03 & 5.9 & 0.0 & 5.93 & 0.56 & 47.88 & 41.98 & - & - \\
		2009 & 2 & K-Means & x & 9.97 & 2.44 & 0.0 & 2.91 & 0.84 & 46.36 & 48.8 & - & - \\
		\hline
		1989 & 2 & Spectral & x & 15.46 & 1.29 & 0.0 & 3.97 & 5.72 & 44.05 & 42.76 & - & - \\
		1995 & 2 & Spectral & x & 14.2 & 3.0 & 0.0 & 4.41 & 2.26 & 44.11 & 47.11 & - & - \\
		2009 & 2 & Spectral & x & 11.68 & 0.39 & 0.0 & 2.25 & 9.3 & 46.64 & 46.25 & - & - \\
		\midrule
		1989 & 3 & K-Means & & 12.11 & 4.24 & 0.19 & 4.51 & 1.65 & 41.64 & 47.44 & 48.0 & - \\
		1995 & 3 & K-Means & & 12.47 & 5.15 & 1.29 & 5.23 & 0.96 & 46.35 & 49.11 & 41.38 & - \\
		2009 & 3 & K-Means & & 10.06 & 2.61 & 0.0 & 2.89 & 0.94 & 48.59 & 45.16 & 49.08 & - \\
		\hline 
		1989 & 3 & Spectral & & 14.81 & 4.66 & 0.01 & 4.83 & 1.4 & 47.46 & 41.69 & 40.47 & - \\
		1995 & 3 & Spectral & & 15.56 & 2.24 & 0.0 & 2.91 & 3.86 & 45.04 & 41.82 & 41.68 & - \\
		2009 & 3 & Spectral & & 11.11 & 2.51 & 0.0 & 2.57 & 1.09 & 48.81 & 45.04 & 46.4 & - \\
		\hline 
		1989 & 3 & K-Means & x & 11.78 & 4.95 & 0.51 & 5.01 & 0.88 & 46.66 & 48.78 & 41.35 & - \\
		1995 & 3 & K-Means & x & 11.99 & 4.45 & 0.0 & 4.83 & 1.08 & 47.42 & 41.86 & 48.54 & - \\
		2009 & 3 & K-Means & x & 10.17 & 2.88 & 0.43 & 3.03 & 1.17 & 48.47 & 45.11 & 49.43 & - \\
		\hline 
		1989 & 3 & Spectral & x & 12.8 & 4.32 & 0.01 & 4.71 & 2.33 & 41.18 & 46.88 & 47.66 & - \\
		1995 & 3 & Spectral & x & 13.93 & 5.11 & 0.0 & 5.42 & 1.48 & 44.16 & 41.73 & 49.39 & - \\
		2009 & 3 & Spectral & x & 10.97 & 2.1 & 0.0 & 2.47 & 1.43 & 45.42 & 46.36 & 48.57 & - \\
		\midrule
		1989 & 4 & K-Means & & 12.86 & 4.38 & 0.14 & 4.58 & 1.77 & 48.51 & 41.28 & 42.44 & 47.05 \\
		1995 & 4 & K-Means & & 15.09 & 3.38 & 0.0 & 3.73 & 1.39 & 42.62 & 48.68 & 49.06 & 47.71 \\
		2009 & 4 & K-Means & & 10.73 & 2.8 & 0.45 & 2.97 & 1.57 & 47.6 & 44.85 & 50.31 & 48.05 \\
		\hline 
		1989 & 4 & Spectral & & 12.23 & 4.33 & 0.21 & 4.39 & 1.6 & 41.25 & 47.21 & 49.12 & 44.86 \\
		1995 & 4 & Spectral & & 15.09 & 3.78 & 0.0 & 4.32 & 3.54 & 44.62 & 41.73 & 41.73 & 48.32 \\
		2009 & 4 & Spectral & & 10.73 & 2.46 & 0.0 & 2.76 & 1.72 & 48.91 & 45.32 & 45.0 & 48.37 \\
		\hline 
		1989 & 4 & K-Means & x & 14.65 & 4.33 & 1.23 & 4.35 & 0.83 & 41.21 & 47.53 & 39.79 & 43.96 \\
		1995 & 4 & K-Means & x & 11.93 & 4.14 & 0.77 & 4.37 & 1.75 & 41.41 & 47.73 & 45.97 & 49.11 \\
		2009 & 4 & K-Means & x & 12.22 & 2.14 & 0.0 & 2.28 & 0.65 & 45.35 & 48.75 & 44.57 & 45.04 \\
		\hline 
		1989 & 4 & Spectral & x & 12.36 & 3.53 & 0.25 & 4.12 & 3.85 & 46.28 & 47.44 & 40.9 & 47.57 \\
		1995 & 4 & Spectral & x & 15.73 & 1.95 & 0.0 & 2.52 & 2.81 & 45.3 & 41.74 & 41.55 & 41.47 \\
		2009 & 4 & Spectral & x & 11.59 & 2.04 & 0.0 & 2.19 & 1.7 & 48.76 & 44.93 & 46.35 & 45.58 \\
	\end{tabular}
	\caption{$\sigma$ is the mean of all cluster price standard deviations, $\Delta_\mu$ is the average distance between cluster mean prices, $\Delta_{\text{min}}$ is the mean nearest neighbour cluster distance, $\Delta_{\text{avg}}$ denotes the average distance between all points situated in different clusters, and $DB$ is the Davies-Bouldin index. $DB$ has values in $[0, \infty)$ and measures the average similarity of every cluster with its most similar cluster. A low $DB$ index indicates a good clustering. $\mu_i$ denotes the mean price of cluster $i$. All metrics were computed considering the yearly average price for each node.}
	\label{tab:1}
\end{table}

	\section{Clusterings visualization based on Rand indices} \label{app:rand-visualization}

\begin{figure}[!htp]
    \centering
    \includegraphics[width=.9\textwidth]{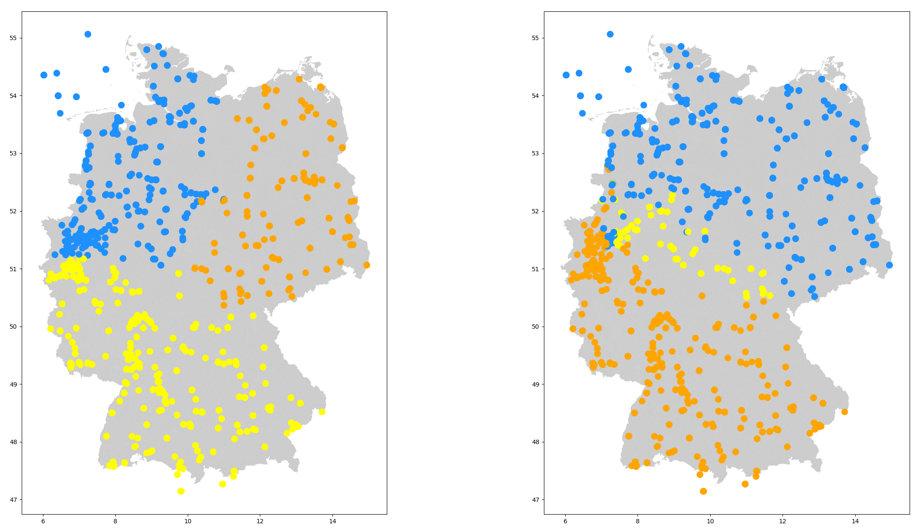} \\
    \caption{RI = 0.44 - left: D3 (spectral); right: Spectral Clustering based on prices; $k = 3$, 2009}
    \label{fig:rand-vis-2-clusters-kmeans}
\end{figure}

\begin{figure}[!htp]
    \centering
    \includegraphics[width=.9\textwidth]{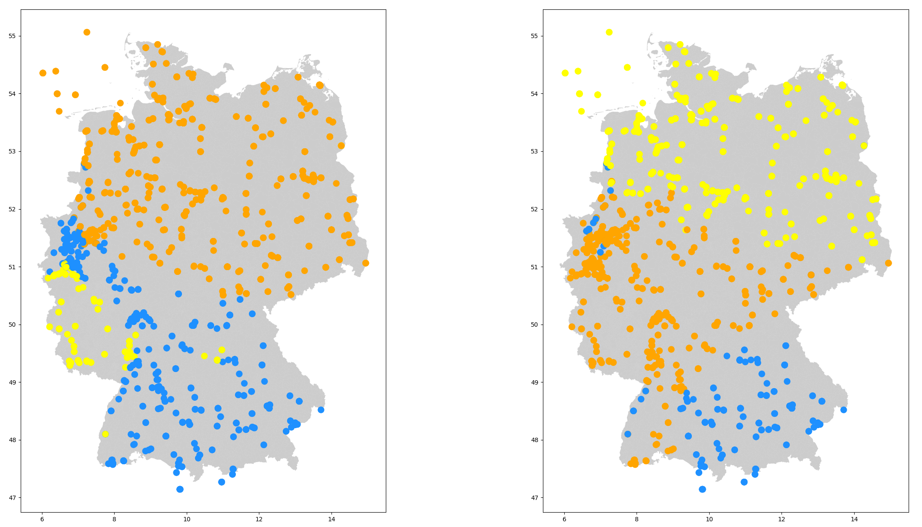} \\    
    \caption{RI = 0.69 - K-Means based on prices; $k = 3$; left: 1989, right: 1995}
    \label{fig:rand-vis-3-clusters-kmeans}
\end{figure}

\begin{figure}[!htp]
    \centering
    \includegraphics[width=.9\textwidth]{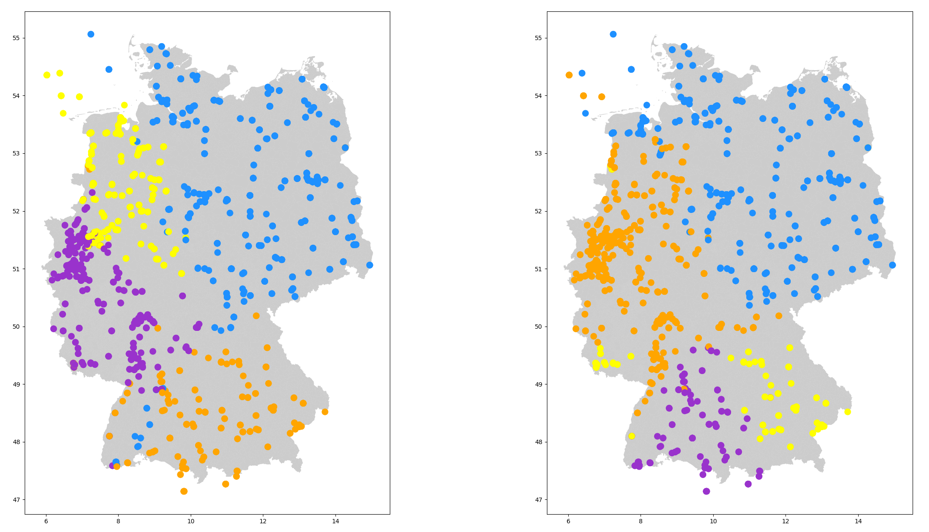} \\
    \caption{RI = 0.83 - K-Means based on prices; $k = 4$; left: 1989, right: 2009}
    \label{fig:rand-vis-4-clusters-kmeans}
\end{figure}

\begin{figure}[!htp]
    \centering
    \includegraphics[width=.9\textwidth]{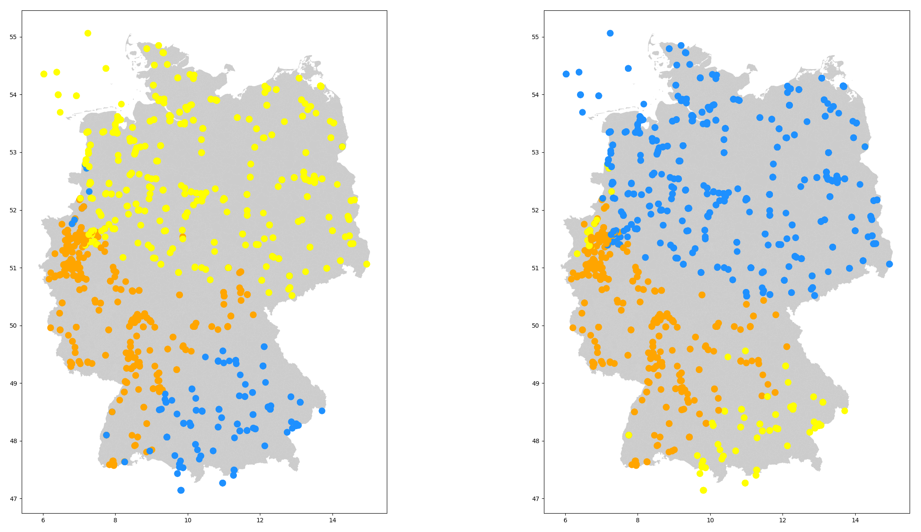} \\
    \caption{RI = 0.93 - K-means based on prices and geographic coordinates; $k = 3$; left: 1989, right: 2009}
    \label{fig:rand-vis-4-clusters-sc}
\end{figure}
	\section{Single-day clustering analysis} \label{app:one-day-analysis}

The time granularity of the prices included in the clustering computation influences the stability of the resulting clusterings. 
In what follows, we analyze the stability of the clusterings identified given the prices corresponding to a single day.
We select the \textcolor{black}{scenario} day of January 22, 1989, and compute clusterings using K-Means and Spectral Clustering.
As clustering feature(s), we consider (1) the time series of prices of a node (12 hours, i.e., 12 features) and (2) the time series of prices of a node and its geographic coordinates.

Table \ref{tab:one-day} shows the values of the inter- and intra-cluster similarity metrics introduced in Section \ref{sec:intra-inter-similarity}.
We observe the average distances between cluster mean prices ($\Delta_\mu$) and the average distances between all points situated in different clusters ($\Delta_{\text{avg}}$) are considerably lower compared to the values shown in Table \ref{tab:1}, indicating that the clusters are similar to each other and thus not well-defined.
We recall that the results illustrated in Table \ref{tab:1} correspond to clusterings obtained considering the annual time series of nodal prices as clustering features.
With time series of smaller length, clusterings appear to be less stable.

\begin{table}[!htp]
	\centering
	\begin{tabular}{c|c|c|c|c|c|c|c|c|c|c|c}
		$k$ & Alg. & Loc. & $\sigma$ & $\Delta_\mu$ & $\Delta_\text{min}$ & $\Delta_\text{avg}$ & $DB$ & $\mu_1$ & $\mu_2$ & $\mu_3$ & $\mu_4$ \\
		\hline 
		2 & K-Means &  & 12.78 & 0.81 & 0.0 & 0.8 & 0.94 & 56.31 & 57.12 & - & - \\
        2 & K-Means & x & 16.17 & 0.39 & 0.0 & 0.56 & 2.03 & 56.35 & 56.74 & - & - \\
        2 & Spectral &  & 13.23 & 0.76 & 0.0 & 0.78 & 1.01 & 56.32 & 57.08 & - & - \\
        2 & Spectral & x & 11.28 & 0.33 & 0.0 & 0.39 & 0.73 & 56.43 & 56.1 & - & - \\
        \hline
        3 & K-Means &  & 11.16 & 0.73 & 0.09 & 0.75 & 0.83 & 56.34 & 56.07 & 57.17 & - \\
        3 & K-Means & x & 10.38 & 0.23 & 0.0 & 0.34 & 1.09 & 56.43 & 56.4 & 56.08 & - \\
        3 & Spectral &  & 12.62 & 0.77 & 0.0 & 0.78 & 1.04 & 56.36 & 57.13 & 55.97 & - \\
        3 & Spectral & x & 13.44 & 0.47 & 0.0 & 0.55 & 1.71 & 56.39 & 56.78 & 56.07 & - \\
        \hline
        4 & K-Means &  & 12.88 & 0.57 & 0.08 & 0.57 & 1.11 & 56.61 & 56.13 & 57.23 & 56.47 \\
        4 & K-Means & x & 12.94 & 0.3 & 0.0 & 0.39 & 5.21 & 56.68 & 56.08 & 56.35 & 56.33 \\
        4 & Spectral &  & 13.02 & 0.67 & 0.0 & 0.68 & 0.94 & 56.39 & 57.2 & 55.97 & 56.08 \\
        4 & Spectral & x & 14.52 & 0.16 & 0.0 & 0.27 & 3.11 & 56.29 & 56.14 & 56.44 & 56.35 \\				
	\end{tabular}
	\caption{$\sigma$ is the mean of all cluster price standard deviations, $\Delta_\mu$ is the average distance between cluster mean prices, $\Delta_{\text{min}}$ is the mean nearest neighbour cluster distance, $\Delta_{\text{avg}}$ denotes the average distance between all points situated in different clusters, and $DB$ is the Davies-Bouldin index. $DB$ has values in $[0, \infty)$ and measures the average similarity of every cluster with its most similar cluster. A low $DB$ index indicates a good clustering. $\mu_i$ denotes the mean price of cluster $i$. All metrics were computed considering the yearly average price for each node.}
	\label{tab:one-day}
\end{table}

\end{appendices}

\end{document}